\documentclass[a4paper,11pt]{article}
\pdfoutput=1 

\usepackage{jheppub} 

\usepackage{graphicx,booktabs}
\usepackage{float}
\usepackage{dcolumn}
\usepackage{bm}
\usepackage{hyperref}  
\usepackage{multirow}
\usepackage{longtable}
\usepackage{xspace}
\usepackage{color}
\usepackage{epsfig}
\usepackage{subfig}
\usepackage{url}
\usepackage{calrsfs}
\usepackage{dsfont}
\usepackage{amssymb}
\usepackage{amsmath}
\usepackage{amsthm}
\usepackage{textcomp}

\DeclareMathAlphabet{\pazocal}{OMS}{zplm}{m}{n}
\def\jpsi{{J/\psi}}

\def\p0{{\bigl.^3\hspace{-1mm}P^{[8]}_0}}

\def\bqa{\begin{eqnarray}}
\def\eqa{\end{eqnarray}}
\def\bc{\begin{center}}
\def\bc{\end{center}}

\def\be{\begin{equation}}
\def\ee{\end{equation}}
\def\bea{\begin{eqnarray}}
\def\eea{\end{eqnarray}}

\providecommand{\MGaMC}{{\sc\small MadGraph5\_aMC@NLO}}
\providecommand{\MGshort}{{\sc\small MG5\_aMC}}
\providecommand{\pythia}{{\sc pythia}}
\newcommand{\mcfm}{{\sc mcfm}}
\newcommand{\toppp}{\ttt{Top++}}

\newcommand{\sqrts}{\sqrt{s}}
\newcommand{\LumiInt}{\mathcal{L}_{\rm \tiny{int}}}
\newcommand{\ttbar}    {t\overline{t}}
\newcommand{\bbbar}    {b\overline{b}}
\newcommand{\ccbar}    {c\overline{c}}
\newcommand{\qqbar}    {q\overline{q}}
\newcommand{\qbar}    {{\overline{q}}}
\newcommand{\mt}{m_{t}}

\newcommand{\lqcd}{\Lambda_{_{\rm QCD}}}
\newcommand{\pt}{p_{_{\mathrm{T}}}}
\newcommand{\alphas}    {\alpha_{\rm s}}

\newcommand{\HT}{\ensuremath{{H\!\!\!/}_{\mathrm{T}}}}

\newcommand{\cm}{c.m.\@\xspace}

\newcommand{\ie}{i.e.\@\xspace}
\newcommand{\eg}{e.g.\@\xspace}

\def\ttt#1{\texttt{\small #1}}

\title{Rare two-body decays of the top quark into a bottom meson plus an up or charm quark}

\author[a]{David d'Enterria}
\author[b]{\! and Hua-Sheng Shao}

\affiliation[a]{CERN, EP Department, CH-1211 Geneva 23, Switzerland}
\affiliation[b]{Laboratoire de Physique Th\'eorique et Hautes Energies (LPTHE), UMR 7589, Sorbonne Universit\'e et CNRS, 4 place Jussieu, 75252 Paris Cedex 05, France}

\emailAdd{david.d'enterria@cern.ch}
\emailAdd{huasheng.shao@lpthe.jussieu.fr}

\abstract{Rare two-body decays of the top quark into a neutral bottom-quark meson plus an up- or charm-quark: $t\to {\overline B}^0+ u, c$; $t\to {\overline B}^0_{s}+ c,u$; and $t \to \Upsilon(nS)+\, c,u$, are studied for the first time. The corresponding partials widths are computed at leading order in the non-relativistic QCD framework. The sums of all two-body branching ratios amount to $\mathcal{B}(t \to {\overline B}^0+\, {\rm jet}) \approx \mathcal{B}(t \to {\overline B}^0_{s}+\, {\rm jet}) \approx 4.2\cdot 10^{-5}$ and $\mathcal{B}(t \to \Upsilon(nS)+\, {\rm jet}) \approx 2\cdot 10^{-9}$, respectively. The feasibility to observe the $t\to {\overline B}^0_{(s)}+{\rm jet}$ decay is estimated in top-pair events produced in proton-proton collisions at $\sqrts = 14,\,100$~TeV at the LHC and FCC, respectively. Combining many exclusive hadronic ${\overline B}^0_{(s)}$ decays, with $\jpsi$ or $ D^{0,\pm}$ final states, about 50 (16\,000) events are expected in 3~(20)~ab$^{-1}$ of integrated luminosity at the LHC (FCC), after typical selection criteria, acceptance, and efficiency losses. An observation of the two-body top-quark decay can also be achieved in the interesting $t\to b(\rm{jet})+c(\rm{jet})$ dijet final state, where the ${\overline B}^0_{(s)}$ decay products are reconstructed as a jet, with 5\,300 and 1.4 million signal events above backgrounds expected after selection criteria at the LHC and FCC, respectively. Such unique final states provide a new direct method to precisely measure the top-quark mass via simple 2-body invariant mass analyses.}

\keywords{top quark, rare decays, $\overline{B}^0$ and $\overline{B}^0_s$ mesons, bottomonia, heavy-flavour jets, QCD, top-quark mass, proton-proton collisions, LHC, FCC}

\begin{document}

\maketitle
\flushbottom


\section{Introduction}

\begin{figure}[t!]
\centering
\includegraphics[width=0.67\textwidth]{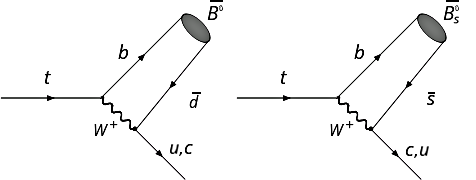}
\includegraphics[width=0.32\textwidth]{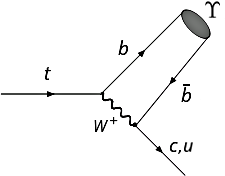}
\caption{\label{fig:feynman}Tree-level Feynman diagrams for the two-body top-quark decays $t\to {\overline B}^0+u,c$ (left), $t\to {\overline B}^0_{s}+c,u$ (center), and $t\to \Upsilon+c,u$ (right). The quark ordering in the labels ($u, c$ or $c, u$) indicates the hierarchy of most probable decays, as computed hereafter.}
\end{figure}

The top quark decays with $\sim$100\% probability into a $W^\pm$ boson plus a $b$-quark, $t\to W b$. The $W^\pm$ boson itself decays in about 2/3 of the cases into hadrons, via the Cabibbo--Kobayashi--Maskawa (CKM) enhanced channels\footnote{Hereafter, unless explicitly stated otherwise, the $t\to W^+b\to\overline{B}^0_{(s)}+X$ decays quoted by default are meant to represent also the equivalent $\overline{t}\to W^-b\to B^0_{(s)}+X$ charge-conjugate decays.} $W^+ \to u\,\overline{d},\; c\,\overline{s}$ (proportional to $ V_{ud,cs} = 0.974$), and much less so via the CKM-suppressed $W^+ \to u\,\overline{s},\; c\,\overline{d}$ (with $ V_{us,cd} = 0.225$) and $W^+ \to c\,\overline{b},\; u\,\overline{b}$ (with $ V_{cb}, V_{ub}\approx 0.04, 0.004$) modes~\cite{PDG}. In this work, we consider the process where one of the down-type quarks $(d, s, b)$ from the $W^\pm$ decay subsequently recombines with the $b$-quark coming directly from the top decay to form a ${\overline B}^0$- or $\Upsilon$-meson, thereby leading to the two-body final-states shown in Fig.~\ref{fig:feynman}. To our knowledge, such decays have never been studied in the literature so far. If such top-quark decays have an experimentally visible branching ratio, they can be used \eg\ to derive an independent value of the top mass ($\mt$) through a simple invariant mass analysis of ${\overline B}^0_{(s)}$- (or $\Upsilon$)-plus-jet pairs around $\mt$. The top mass is one of the key parameters of the Standard Model (SM), chiefly affecting theoretical predictions of Higgs boson properties and searches for new physics beyond the SM (BSM), as well as playing a leading role in the stability of the electroweak vacuum at asymptotic energies~\cite{Alekhin:2012py}. Having at hand different methods to determine $\mt$, and thereby precisely derive its value, is essential for testing the overall SM consistency, reducing parametric uncertainties in the extraction of other various key SM parameters, and constraining BSM models through precision electroweak fits.

Rare top-quark decays can be measured by exploiting the very large top-quark data samples expected to be collected in proton-proton (pp) collisions at centre-of-mass (\cm) energies of $\sqrts =$~14~TeV during the high-luminosity phase of the Large Hadron Collider (HL-LHC)~\cite{Apollinari:2017cqg}, as well as at $\sqrts =$~100~TeV at the planned Future Circular Collider (FCC)~\cite{Benedikt:2018csr,Mangano:2016jyj}. At the HL-LHC, the ATLAS and CMS experiments are expected to integrate luminosities of the order of $\LumiInt = 300$~fb$^{-1}$ per year, adding up to 3~ab$^{-1}$ over the projected lifetime of ten years, whereas the FCC-pp plans to integrate 20~ab$^{-1}$ over twenty years of operation. For the $\sim$1 (35)~nb inclusive top-pair cross section expected at $\sqrts =$~14 (100)~TeV~\cite{Azzi:2017iwa,Mangano:2016jyj}, the experiments would thereby be able to analyse about $6\cdot10^{9}$ ($1.5\cdot10^{12}$) top-plus-antitop quark decays for rare searches such as the one under consideration here.\\

In this paper, we compute first the branching ratios for the two-body decays shown in Fig.~\ref{fig:feynman} making use of the non-relativistic QCD framework to describe the $(b\overline{d}), (b\overline{s})$, and $(b\overline{b})$ bound-state formation from the top-quark decay products (Section~\ref{sec:theory}). A detailed study of the experimental feasibility of measuring such rare two-body decays at the LHC and FCC is described in Section~\ref{sec:measure}, in many different ${\overline B}^0_{(s)}$ decay modes. Section~\ref{sec:t_bc} discusses the observation of the top-quark 2-body decay into bottom+charm jets, including estimates of the expected precision on $\mt$ based on $b+c$ dijet invariant mass analyses. The paper ends with a summary of the main results in Section~\ref{sec:summary}.

\section{Theoretical calculations}
\label{sec:theory}

The production of a $B$ meson from a $b$-quark, besides the well-known $b\to B$ parton fragmentation process, can also proceed through a quark \textit{recombination} mechanism~\cite{Braaten:2001bf} whereby a lighter $d$ or $s$ antiquark from the same hard scattering that produces the bottom quark combines with the latter to form a $\overline{B}^0 \equiv (b\overline{d})$ or $\overline{B}^0_s \equiv (b\overline{s})$ 
bound state. In such cases, the momentum of the $B$ meson arises from the combination of the down or strange plus bottom quark momenta, where both quarks are moving along the same direction. 

To better understand the process, it is informative to consider first the kinematics of such a two-body decay in more detail. In order to form a bound state $B$, the four-momenta $P^\mu$ of the $B$ meson and of the $b$ and $\qbar$ quarks in Fig.~\ref{fig:feynman} have to satisfy:
\begin{equation*}
P^\mu_b = \frac{m_b}{m_b+m_\qbar}P^\mu_B, \quad P^\mu_\qbar = \frac{m_\qbar}{m_b+m_\qbar}P^\mu_B\,.
\end{equation*}
From this, it follows that the invariant mass of the quarks from the $W\to q'\qbar$ decay
must be of the order of $\mathcal{O}(\mt\sqrt{m_\qbar/m_b})\approx 45$~GeV, 
\ie\ the $W$ boson is off-shell with $|m_{W^*\to q'\qbar} - m_W| \gg \Gamma_W\approx 2$~GeV.  
In the rest frame of the top quark, the leading-order (LO) kinematics of the two-body decay leads to the following typical momenta of the $\overline{B}^0_{(s)}$ meson and back-to-back quark $q$:
\begin{eqnarray}
p_{B}&=&\{\varepsilon(\mt,m_{B},m_q),\pi(\mt,m_B,m_q)\overrightarrow{e}\},\nonumber\\
p_{q}&=&\{\varepsilon(\mt,m_q,m_B),-\pi(\mt,m_B,m_q)\overrightarrow{e}\},
\end{eqnarray}
where $\overrightarrow{e}$ is a unity three-dimensional vector $|\overrightarrow{e}|=1$ and
\begin{eqnarray}
\pi(\mt,m_B,m_q)&=&\frac{1}{2\mt}\sqrt{\mt^4+m_B^4+m_q^4-2\mt^2m_B^2-2\mt^2m_q^2-2m_B^2m_q^2} \approx \frac{\mt}{2}\nonumber\\
\varepsilon(\mt,m_B,m_q)&=&\frac{\mt^2+m_B^2-m_q^2}{2\mt} \approx \frac{\mt}{2}.
\end{eqnarray}
With the replacement $q\rightarrow W$ 
in the expressions above, one can see that the bottom quark shares more momentum in the two-body decays of Fig.~\ref{fig:feynman} than in the standard top-quark final state where the $W$ boson 
is on-shell. 
In this latter case, the energies of the $W$-decay quarks are typically 
$(\mt^2+m_W^2)/(4\mt)\approx 50$~GeV, whereas in the two-body decay 
the up-type quark from the $t\to \overline{B}^0\,q$ decay carries an energy of the order of $\mt/2$ and is thereby more boosted than in the common $t\to b\,W \to b\,q'\qbar$ channel.\\

For the formation of a bound state of a heavy quark and an antiquark, the non-relativistic quantum chromodynamics (NRQCD) approach~\cite{Bodwin:1994jh}, in which the quark interactions are organized in an expansion in $v$ (the typical relative velocity of the quarks inside the meson state) provides a useful theoretical framework. The relevant partial width calculations are quite analogous to those for $B_c^\pm$ production from final-state $b$ and $c$ quarks in NRQCD~\cite{Braaten:2001bf,Chang:2007si}. For a decay process $a\to \overline{B}+X$, the partial width $\Gamma$ reads
\begin{eqnarray}
d\Gamma[a\to \overline{B}+X]&=&\sum_{n}{d\hat{\Gamma}[a\to (b\overline{q})_n+X]\;\rho[(b\overline{q})_n\to \overline{B}]},
\end{eqnarray}
where $(b\overline{q})_n$ represents a given Fock state $n$ of the bottom $b$ and light-$\overline{q}$ quarks, and the factor $\rho[(b\overline{q})_n\to \overline{B}]$ is a non-perturbative probability for $(b\overline{q})_n$ to evolve into the $\overline{B}$ meson. However, unlike in the case of $B_c^\pm$ production, the non-perturbative transition $(b\overline{q})_n\to \overline{B}$ does not follow a definite velocity power-counting because the relative velocity of the light $\overline{d}$ or $\overline{s}$ quarks in the rest frame of the $B$ meson is not small. Therefore, Fock state $(b\overline{q})_n$ contributions to the $\overline{B}$ meson with different quantum numbers (\eg\ colour or angular momentum) are not necessarily suppressed.

In Ref.~\cite{Braaten:2001bf}, it was argued that one only needs to consider Fock states with the $S$-wave orbital angular momentum, because higher orbital 
states are suppressed by at least a factor of $m_b/\lqcd\approx 
20$. In this case, the only relevant states are  $n=\!^1S_0^{[1]},^3S_1^{[1]},^1S_0^{[8]},^3S_1^{[8]}$, where we use the spectroscopic notation $^{2s+1}L_{J}^{[c]}$ with $s$ being spin, $L$ being orbital angular momentum, $J$ being total angular momentum, and $c = 1, 8$ being the colour singlet and octet representations, respectively. 
By using the heavy-quark spin symmetry, we can further reduce the non-perturbative transition probabilities $\rho$, from four to two, \ie
\begin{eqnarray}
\rho_1^{\overline{B}}&\equiv& \rho[(b\overline{q})_{^1S_{0}^{[1]}}\to \overline{B}], \nonumber\\
\rho_8^{\overline{B}}&\equiv& \rho[(b\overline{q})_{^1S_{0}^{[8]}}\to \overline{B}],\\
3\rho_{1,8}^{\overline{B}} & = & \rho[(b\overline{q})_{^3S_{1}^{[1,8]}}\to \overline{B}].\nonumber 
\end{eqnarray}
The $\rho_1$ and $\rho_8$ probabilities can then be derived from the standard long-distance matrix elements (LDMEs), $\langle \mathcal{O}^{\overline{B}}(^{2s+1}L_{J}^{[c]})\rangle$, widely used in quarkonium physics:
\begin{eqnarray}
\label{eq:rho1_rho8}
\langle \mathcal{O}^{\overline{B}}(^1S_0^{[1]})\rangle&=&2N_c\frac{4m_bm_\qbar^2}{3}\rho_{1,8}^{\overline{B}},\nonumber\\
\langle \mathcal{O}^{\overline{B}}(^1S_0^{[8]})\rangle&=&\left(N_c^2-1\right)\frac{4m_bm_\qbar^2}{3}\rho_{1,8}^{\overline{B}},\\
\langle \mathcal{O}^{\overline{B}}(^3S_1^{[1,8]})\rangle&=&3\langle \mathcal{O}^{\overline{B}}(^1S_0^{[1,8]})\rangle,\nonumber
\end{eqnarray}
where $m_b$ is the bottom quark mass, $m_\qbar = m_{\bar d}, m_{\bar s}\approx 0.3$~GeV are the constituent light quark masses, and $N_c = 3$ the number of colours.

So far the discussion focused on the top decays into $\overline{B}^{0}_{(s)}+$\,jets. In the case of top decays leading to bottomonium states plus a jet, $t\to \Upsilon(nS)+c,u$ (Fig.~\ref{fig:feynman} right), the two bottom quarks from the $W$ and $t$ decays recombine with relative velocity $v\simeq \sqrt{0.1}$ (as given from the $\Upsilon(nS)$ binding energies), the NRQCD velocity scaling rule is well applicable, and various orbital states can contribute. However, as discussed below, such a decay mode is reduced by at least a factor of $|V_{cb,ub}|^2/|V_{ud,cs}|^2\approx 10^{-4}$ compared to the $\overline{B}^0_{(s)}+\rm{jet}$ ones, mostly due to the already very-suppressed decays of $W$ bosons into bottom quarks.

\begin{table*}
\centering
\caption{Numerical values of the SM parameters used in the calculations.\label{tab:params}}
\begin{tabular}{cl|cl}
\toprule
Masses (and top width)& & Other SM parameters & \\\midrule
$m_d$ & 0.3~GeV         & $\sin{\theta_w}$ & $0.471423$ \\
$m_s$ & 0.3~GeV         & $\alpha^{-1}$ & $132.0$  \\
$m_c$ & $1.5$~GeV       & $\alpha_s$    & 0.1180 \\
$m_b$ & $4.75$~GeV      & $|V_{tb}|$ & $0.99$ \\
$\mt$ & $173.0$~GeV     & $|V_{ud,cs}|$ & $0.974$ \\
$m_W$ & $80.419$~GeV    & $|V_{us,cd}|$ & $0.225$ \\
      &                 & $|V_{cb}|$ & $0.04$  \\
$\Gamma_t$ & $1.35$~GeV & $|V_{ub}|$ & $0.004$ \\
\bottomrule
\end{tabular}
\end{table*}

\subsection{Top decays into a \texorpdfstring{$\overline{B}^0_{(s)}$}{B0(s)}-meson plus a \texorpdfstring{$u$}{u}- or \texorpdfstring{$c$}{c}-quark}

We compute first the top decay modes shown in the left and center plots of Fig.~\ref{fig:feynman}. Based on the NRQCD recombination framework described above, the LO (in the QCD coupling $\alpha_s$) partial widths can be obtained from the following expressions
\begin{eqnarray}
\label{eq:Gamma_Bjet}
\Gamma(t\to \overline{B}^0+q+X)&=&\left(\rho_1^{\overline{B}}+8\rho_8^{\overline{B}}\right)|V_{qd}|^2\frac{\alpha^2 \pi}{9}\frac{m_{\bar{d}}^2(\mt^2-m_b^2)^2(\mt^2+m_b^2)}{\mt^3m_W^4\sin^4{\theta_w}}\left[1+\mathcal{O}\left(\frac{m_q}{m_b}\right)\right],\nonumber\\
\Gamma(t\to \overline{B}_{s}^0+q+X)&=&\frac{|V_{qs}|^2}{|V_{qd}|^2}\frac{m_{\bar{s}}^2}{m_{\bar{d}}^2}\Gamma(t\to \overline{B}^0+q+X)\,,
\end{eqnarray}
with the various input parameters listed in Table~\ref{tab:params}.
Here, we write the generic $t\to \overline{B}^0_{(s)}+q+X$ decay that includes not only exclusive two-body $t\to \overline{B}^0_{(s)}+q$ decays, but also semi-inclusive ones from colour octet configurations, those with $c=8$ in Eq.~(\ref{eq:rho1_rho8}), where $X$ is a soft gluon radiated
in the transition into the final colourless $B$ hadron. These soft gluons hadronize into low-energy pions or kaons with energies close to the binding energy of the $B$ hadron ($\Delta E \approx 400$ MeV in the rest frame of the initial $(b\overline{q})_n$ Fock state).
To evaluate numerically the expressions of Eq.~(\ref{eq:Gamma_Bjet}), we use the parameters listed in Table~\ref{tab:params}, with the $\rho_{1,8}^{\overline{B}}$ transition probabilities estimated as in Ref.~\cite{Braaten:2001bf} by relating them to the $\rho_{1,8}^{D}$ ones, where the latter are determined from the $D^*$-meson fragmentation function calculation of Ref.~\cite{Berezhnoy:1999rd}. Assuming $\rho_{1,8}^{\overline{B}}=(m_c/m_b)\rho_{1,8}^{D}$, the values $\rho_1^{\overline{B}}=0.4$ and $\rho_8^{\overline{B}}=0.8$ are obtained for $m_{\bar{q}}=0.3$~GeV. These values are relatively badly known, and are the leading source of uncertainty in our theoretical estimates.
Since $|V_{qs}|^2/|V_{qd}|^2\approx 20, 0.05$ for $q=u,c$, one can see from Eq.~(\ref{eq:Gamma_Bjet}) that the $t\to {\overline B}^0+u,\,{\overline B}^0_{s}+c$ decays are enhanced by a factor of $\times20$ compared to the $t\to {\overline B}^0+c,\, {\overline B}^0_{s}+u$ ones, respectively.

The final numerical result for the partial width of a top quark into $\overline{B}^0+u$ (or $\overline{B}_s^0 + c$) in the NRQCD model is $5.4\cdot 10^{-5}$ GeV, which for $\Gamma_t = 1.35$~GeV~\cite{PDG} results into the corresponding top-quark branching ratios of $\mathcal{B}(t\to \overline{B}^0+u, \overline{B}^0_s+c) = \Gamma(t\to \overline{B}^0_{(s)}+u \,(c))/\Gamma_{\rm t} = 4\cdot 10^{-5}$ and $\mathcal{B}(t\to \overline{B}^0+c, \overline{B}^0_s+u) = 2.1\cdot 10^{-6}$. We assign to those numerical results theoretical uncertainties of the order of 100\%, dominated by the non-perturbative transition probabilities $\rho_{1,8}^{\overline{B}}$. The sum of all such two-body decay modes has a total branching fraction of $\mathcal{B}(t\to {\overline B}^0_{(s)}+{\rm jet}) = 8.4\cdot 10^{-5}$. The first result of our study is, therefore, that top quarks have a small but non-negligible (1-in-12\,000) probability to decay into the 2-body states, $t\to {\overline B}^0+{ u, c}$ and $t\to {\overline B}^0_{s}+{ c,u}$, plotted in the left and center panels of Fig.~\ref{fig:feynman}. All individual partial decay widths and branching fractions computed for each one of the four channels are listed in Table~\ref{tab:results}.

\begin{table*}[htpb!]
\centering
\caption{Partial widths and branching fractions of top-quark two-body decays into a $\overline{B}^0_{(s)}$ meson plus a jet, computed here. The branching fractions assume a total top width of $\Gamma_t = 1.35$~GeV~\cite{PDG}. (Theoretical uncertainties, of the order of 100\%, not quoted).}
\begin{tabular}{lcc}\toprule
 Decay mode & Partial decay width (GeV) & Branching fraction \\\hline
$t\to \overline{B}^0+u$ & $5.4\cdot 10^{-5}$ & $4.0\cdot 10^{-5}$ \\
$t\to \overline{B}^0+c$ & $2.9\cdot 10^{-6}$ & $2.1\cdot 10^{-6}$ \\\hline
$t\to \overline{B}^0_s+c$ & $5.4\cdot 10^{-5}$ & $4.0\cdot 10^{-5}$ \\
$t\to \overline{B}^0_s+u$ & $2.9\cdot 10^{-6}$ & $2.1\cdot 10^{-6}$ \\\hline
$t\to {\overline B}^0_{(s)}+{\rm jet}$ (total) & $1.14\cdot10^{-4}$ & $8.4\cdot 10^{-5}$\\\bottomrule
\end{tabular}
\label{tab:results} 
\end{table*}

\subsection{Top decays into an  \texorpdfstring{$\Upsilon(nS)$}{Ups(nS)}-meson plus a \texorpdfstring{$u$}{u}- or \texorpdfstring{$c$}{c}-quark}

Following the theoretical approach described above, we can similarly calculate the CKM-suppressed $t\to \Upsilon(nS)+c,u$ decay (Fig.~\ref{fig:feynman} right). In this process, the NRQCD velocity scaling rule is well applicable since the relative velocity between the two bottom quarks is $v\simeq \sqrt{0.1}$. We note first that the $t\to \Upsilon(nS)+u$ decay is suppressed by a factor of $|V_{ub}|^2/|V_{cb}|^2\approx 10^{-2}$ compared to the $t\to \Upsilon(nS)+c$ one, and therefore we will just focus on the latter hereafter. The LO expressions for producing a bottomonium $H_{b\overline{b}}$ state, neglecting the subleading terms suppressed by factors of 
$m_b^2/m_W^2$ and/or $m_q^2/m_b^2$, are 
\begin{eqnarray}
\Gamma(t\to \left(H_{b\overline{b}}\right)_{^3S_1^{[1]}}+q)&=&\frac{\langle \mathcal{O}^{H_{b\overline{b}}}(^3S_1^{[1]})\rangle}{m_b^3}\frac{\Gamma_0(t\to \left(b\overline{b}\right)_S+q)}{216},\nonumber\\
\Gamma(t\to \left(H_{b\overline{b}}\right)_{^3S_1^{[8]}}+q)&=&\frac{\langle \mathcal{O}^{H_{b\overline{b}}}(^3S_1^{[8]})\rangle}{m_b^3}\frac{\Gamma_0(t\to \left(b\overline{b}\right)_S+q)}{36},\\
\Gamma(t\to \left(H_{b\overline{b}}\right)_{^1S_0^{[8]}}+q)&=&\frac{\langle \mathcal{O}^{H_{b\overline{b}}}(^1S_0^{[8]})\rangle}{m_b^3}\frac{\Gamma_0(t\to \left(b\overline{b}\right)_S+q)}{12},\nonumber
\end{eqnarray}
for the $S$-wave states, and
\begin{eqnarray}
\Gamma(t\to \left(H_{b\overline{b}}\right)_{^3P_{0}^{[8]}}+q)&=&\frac{\langle \mathcal{O}^{H_{b\overline{b}}}(^3P_0^{[8]})\rangle}{m_b^5}\frac{\Gamma_0(t\to \left(b\overline{b}\right)_P+q)}{36},\nonumber\\
\Gamma(t\to \left(H_{b\overline{b}}\right)_{^3P_{1}^{[8]}}+q)&=&\frac{\langle \mathcal{O}^{H_{b\overline{b}}}(^3P_1^{[8]})\rangle}{m_b^5}\frac{\Gamma_0(t\to \left(b\overline{b}\right)_P+q)}{18}\left(1-2\frac{m_W^2}{\mt^2}\right)^2,\nonumber\\
\Gamma(t\to \left(H_{b\overline{b}}\right)_{^3P_{2}^{[8]}}+q)&=&\frac{\langle \mathcal{O}^{H_{b\overline{b}}}(^3P_2^{[8]})\rangle}{m_b^5}\frac{\Gamma_0(t\to \left(b\overline{b}\right)_P+q)}{90},\nonumber\\
\Gamma(t\to \left(H_{b\overline{b}}\right)_{^3P_{0}^{[1]}}+q)&=&\frac{\langle \mathcal{O}^{H_{b\overline{b}}}(^3P_0^{[1]})\rangle}{m_b^5}\frac{\Gamma_0(t\to \left(b\overline{b}\right)_P+q)}{216},\\ 
\Gamma(t\to \left(H_{b\overline{b}}\right)_{^3P_{1}^{[1]}}+q)&=&\frac{\langle \mathcal{O}^{H_{b\overline{b}}}(^3P_1^{[1]})\rangle}{m_b^5}\frac{\Gamma_0(t\to \left(b\overline{b}\right)_P+q)}{108}\left(1-2\frac{m_W^2}{\mt^2}\right)^2,\nonumber\\
\Gamma(t\to \left(H_{b\overline{b}}\right)_{^3P_{2}^{[1]}}+q)&=&\frac{\langle \mathcal{O}^{H_{b\overline{b}}}(^3P_1^{[2]})\rangle}{m_b^5}\frac{\Gamma_0(t\to \left(b\overline{b}\right)_P+q)}{540},\nonumber
\end{eqnarray}
for the $P$-wave states, where $\langle \mathcal{O}^{H_{b\overline{b}}}(^{2s+1}L_j^{[c]})\rangle$ are the LDMEs encoding the non-perturbative formation of the $\Upsilon(nS)$ bound state, and we define
\begin{eqnarray}
\Gamma_0(t\to \left(b\overline{b}\right)_S+q)&\equiv & \left|V_{qb}\right|^2\frac{\alpha^2 \pi \mt^3m_b^2}{\left(\mt^2-2m_W^2\right)^2\sin^4{\theta_w}},\nonumber\\
\Gamma_0(t\to \left(b\overline{b}\right)_P+q)&\equiv& \left|V_{qb}\right|^2\frac{\alpha^2 \pi \mt^7m_b^2}{\left(\mt^2-2m_W^2\right)^4\sin^4{\theta_w}}.
\end{eqnarray}
For $\Upsilon$ mesons, the NRQCD velocity scaling rule suggests that $P$-wave states have the same power counting as the colour-octet $S$-wave channels, which in any case are much smaller than the leading-power counting channel driven by the colour-singlet $S$-wave, $^3S_1^{[1]}$. We include here all orbital channels because their associated LDMEs are well known (which is clearly not the case for the ${\overline B}^0$ mesons considered in the previous section). In our numerical results, we keep all the $m_b$ and $m_c$ dependencies at LO with the parameters listed in Table~\ref{tab:params}. In addition, similar to what has been done in Ref.~\cite{Shao:2016wor}, we consider four different sets of non-perturbative LDMEs for bottomonia production:
\begin{itemize}
\item{Set I:} This set is based on the LDMEs presented in Ref.~\cite{Kramer:2001hh}, with the $\chi_b(3P)$ ones set to zero.
\item{Set II:} This set is based on the bottomonia LDMEs extracted in Ref.~\cite{Sharma:2012dy}, where the $\chi_b(3P)$ contributions are also ignored.
\item{Set III:} Unlike the previous two sets, we used the LDMEs extracted from next-to-leading-order (NLO) analyses. The LDMEs of $\Upsilon(nS),\chi_b(nP),1\le n\le 3$ are from Ref.~\cite{Han:2014kxa}.
\item{Set IV:} A second set of LDMEs based on NLO calculations taken from Ref.~\cite{Feng:2015wka} for bottomonia.
\end{itemize}
The final numerical values of the partial widths for the various bottomonium state decays, after including all feed-down contributions, are listed in Table~\ref{tab:Ywidth}. 

\begin{table*}[htpb!]
\centering
\caption{Partial widths (in units of $10^{-9}$ GeV) of inclusive top-quark decays into $\Upsilon(1S,2S,3S)$ mesons (including feed-down contributions from higher-excited bottomonia decays) plus a quark, for the four LDME sets considered in this work.}
\label{tab:Ywidth} 
\begin{tabular}{ccccc}\toprule 
\rule{0pt}{3ex}
 top decay  & \multicolumn{4}{c}{Partial decay width ($10^{-9}$ GeV)}\\
 (bottomonium state) & Set I & Set II & Set III & Set IV \\[1mm] 
\hline\rule{0pt}{3ex}
 $t\to \Upsilon(1S)+q$ & $1.62$  & $2.01$ & $1.61$ & $1.36$ \\
\rule{0pt}{3ex}
 $t\to \Upsilon(2S)+q$  &  $0.71$ & $0.23$ & $0.67$ & $0.57$ \\
\rule{0pt}{3ex}
 $t\to \Upsilon(3S)+q$  &  $0.51$ & $0.47$ & $0.42$ & $0.37$ \\\hline
\rule{0pt}{3ex}
$t\to \Upsilon(nS)+q$  &  $2.84$ & $2.71$ & $2.70$ & $2.30$ \\
\bottomrule 
\end{tabular}
\end{table*}

The $\Upsilon(1S,2S,3S)+$\,quark decays add up to a partial width around $2.5\cdot 10^{-9}$ GeV (with a small $\sim$10\% theoretical uncertainty given by the span of the results obtained with the different LDME sets listed in Table~\ref{tab:Ywidth}), which is four orders of magnitude smaller than that corresponding to the $\overline{B}_{(s)}^0$ mode in the recombination mechanism (Table~\ref{tab:results}). Such a large difference is explained by the very small probability of the $W$ boson to decay into bottom quarks to start with, $|V_{cb,ub}|^2/|V_{ud,cs}|^2\approx 10^{-3}\rm{-}10^{-5}$, followed by the suppression of $\Upsilon$ colour-octet channels. The individual partial widths and branching fractions computed for the  top decays into bottomonium+jet are listed in Table~\ref{tab:results_Upsilon}.

\begin{table*}[htpb!]
\centering
\caption{Partial widths and branching fractions of top-quark two-body decays into an $\Upsilon$ meson plus a jet computed here. The branching fractions consider a total top-quark width of $\Gamma_t = 1.35$~GeV~\cite{PDG}. (Theoretical uncertainties, of the order of 10\%, not quoted).}
\begin{tabular}{lcc}\toprule
 Decay mode & Partial decay width (GeV) & Branching fraction \\\hline
$t\to \Upsilon(nS)+c$ & $2.5\cdot 10^{-9}$ & $1.9\cdot 10^{-9}$ \\
$t\to \Upsilon(nS)+u$ & $2.5\cdot 10^{-11}$ & $1.9\cdot 10^{-11}$ \\
\bottomrule
\end{tabular}
\label{tab:results_Upsilon} 
\end{table*}

\section{Measurement of \texorpdfstring{$t\to {\overline B}^0_{(s)}+$\,jet}{t to B0+jet} in pp collisions at the LHC and FCC}
\label{sec:measure}

From the computed branching fractions of the rare two-body top decays listed in Tables~\ref{tab:results} and \ref{tab:results_Upsilon}, it is clear that the possibility to experimentally observe any of them is small and should be focused solely on the $\overline{B}^0_{(s)}$-meson--plus--jet final states, as they are at least four orders-of-magnitude more probable than the $\Upsilon$-plus-jet ones. In order to maximize the size of the top-quark data sample, we consider top-quark pair ($\ttbar$) production in pp collisions at $\sqrts =$~14~TeV during the HL-LHC phase, as well as at $\sqrts =$~100~TeV at the FCC. The rare two-body decays under consideration here have branching ratios, $\mathcal{O}(10^{-5})$, of the same order as (or two orders-of-magnitude larger than) those accessible in BSM searches for rare $t\to q\gamma, \, q Z$ decays via flavour-changing neutral currents at the HL-LHC~\cite{Mandrik:2018gud} and FCC~\cite{Mangano:2016jyj}, which also exploit the very large top-quark data samples available at both machines.

In the conservative estimates presented hereafter, we will assume detector acceptances and reconstruction performances typical of the current LHC analyses. At the HL-LHC, the upgraded machine and detectors will function at luminosities $5\rm{-}7.5 \cdot 10^{34}$~cm$^{-2}$s$^{-1}$, corresponding to an average pileup of 140--200 pp collisions per bunch crossing. Advanced pileup mitigation techniques have been therefore designed to keep the top-quark reconstruction under control~\cite{Azzi:2017iwa}, and no significant degradation of the energy scale and resolution of the jets, the missing transverse energy (neutrinos), nor the $b$-jet identification are expected compared to the current ATLAS and CMS performances\footnote{The LHCb detector has unparalleled $\overline{B}^0_{(s)}$ reconstructions capabilities~\cite{LHCb:2011aa,Aaij:2011ac,LHCb:2012ae,Aaij:2013noa,Aaij:2013orb,Aaij:2013zpt,CMS:2014xfa,Aaij:2015fea,Aaij:2011jp,Aaij:2012zka,Aaij:2013dda,Aaij:2013pua,Aaij:2015sqa,Aaij:2018jqv,Aaij:2018rol,Aaij:2019lkm}, albeit with (smaller) forward-only acceptance compared to ATLAS/CMS. Although $\ttbar$ production has been already observed in its dilepton final-state~\cite{Aaij:2015mwa,Aaij:2018imy}, LHCb will integrate about a factor of 10 lower luminosities than ATLAS and CMS in the HL-LHC phase, and the feasibility of this search would thereby require a dedicated analysis beyond this first exploratory study.}. Any unexpected loss in the top-quark reconstruction efficiency due to potentially increased kinematic thresholds (at the trigger or offline analysis levels) will be compensated by the higher $\ttbar$ cross section at 14 TeV compared to the 13-TeV operation so far, as well as by the increased (forward) acceptance and higher granularity of the upgraded detectors, compared to the LHC conditions today. At the FCC, the experiments aim at fully tracking coverage over a large pseudorapidity $|\eta| < 5$ region, leading to $\sim$1.7 times effectively larger acceptance for top-quark detection than currently at the LHC, with a detector granularity also adapted to cope with the expected $\mathcal{O}(200\rm{-}1000)$ pileup collisions.

\subsection{Expected top-quark yields}

The total pp\,$\to\ttbar+X$ cross sections at $\sqrts = 14$ and 100~TeV are computed with $\toppp$ v2.0~\cite{Czakon:2013goa} at next-to-next-to-leading order (NNLO) plus next-to-next-to-leading-log (NNLL) accuracies, using the NNLO 
NNPDF3.0 parton distribution functions (PDF)~\cite{Ball:2014uwa}. 
The code is run with $\rm N_f=5$ flavours, top pole masses set to $\mt=173.0$~GeV, default renormalization and factorization scales set to $\mu_{\textsc{r}}=\mu_{\textsc{f}}=\mt$, and QCD coupling set to $\alphas$~=~0.1180. 
Such a theoretical NNLO+NNLL setup yields theoretical cross sections that are in very good agreement with all the experimental data measured so far in pp collisions at $\sqrts$~=~7,~8,~13~TeV at the LHC~\cite{PDG}. The computed NNLO+NNLL cross sections amount to $\sigma_{\ttbar} = 980 \pm 17\,(\textsc{pdf})\,^{+24}_{-35} (\rm{scale})$~pb at the LHC, and $\sigma_{\ttbar} = 34.80 \pm 1.20\,(\textsc{pdf})\,^{+1.00}_{-1.65} (\rm{scale})$~nb at the FCC. The PDF uncertainties, of the order of $\pm$1.8\% at the LHC and $\pm$3.5\% at the FCC, are obtained from the 100 replicas of the NNPDF3.0 set. The theoretical scale uncertainty, amounting to $+2.5\%,-3.5\%$ at the LHC and $+2.9\%,-4.7\%$ at the FCC, is estimated by modifying $\mu_{\textsc{r}}$ and $\mu_{\textsc{f}}$ within a factor of two with respect to their default value. The computed total cross sections are listed in the first column of Table~\ref{tab:acceptance}. From these cross sections, one would expect a total inclusive number of two-body-decay events of the order of: 
\begin{eqnarray}
N_{\rm t\to \overline{B}^0_{(s)}+j}^{\ttbar, \rm evts}  = \sigma_{\ttbar} \times \LumiInt \times \mathcal{B}(t \to \overline{B}^0_{(s)}+{\rm jet}) & \approx & \nonumber\\ 
\approx 1\,\rm{nb} \times 3\cdot10^{9}\,\rm{nb}^{-1} \times 8.4\cdot10^{-5}  & \approx & 2.5\cdot10^{5}
\label{eq:N_tt_LHC}
\end{eqnarray}
at the HL-LHC, and 
\begin{equation}
N_{\rm t\to \overline{B}^0_{(s)}+j}^{\ttbar,\rm evts} \approx 35\,\rm{nb} \times 2 \cdot 10^{10}\, \rm{nb}^{-1} \times 8.4\cdot10^{-5} \approx 6\cdot 10^{7}
\label{eq:N_tt_FCC}
\end{equation}
at the FCC. The numbers of rare top-quark decays at the FCC is 250 times larger than at the LHC, thanks to $\times$35 and $\times$7 larger cross sections and integrated luminosities, respectively

Of course, the number of such decays actually visible will be smaller after taking into account detector acceptance and reconstruction efficiency losses for the final states under consideration. Depending on the decay of one or both $W$ bosons, the final states for the top-pair process can be divided into three classes~\cite{PDG}:
\begin{itemize}
    \item Fully hadronic: $\ttbar \to W^+ b\,\,W^- \overline{b} \to q'\qbar b \,\,\qbar''q'''\overline{b}$ ($\mathcal{B} = 45.7\%$)
    \item Lepton+jets\footnote{This decay chain includes charge-conjugate mode.}: $\ttbar \to W^+ b\,\,W^- \overline{b} \to q'\qbar b \,\, \ell^- \nu_\ell\overline{b}$ ($\mathcal{B} = 43.8\%$)
    \item Dileptons: $\ttbar \to W^+ b\,\,W^- \overline{b} \to \ell^+ \nu_\ell b \,\, \ell^- \nu_\ell\overline{b}$ ($\mathcal{B} = 10.5\%$)
\end{itemize}
Obviously, the last purely leptonic final-states cannot be used for searches of the rare 2-body hadronic decays considered here, which leaves us with 90\% of the $\ttbar$ cross section, with fully-hadronic and lepton+jets final-states, to be exploited as described next.

\begin{table*}[htbp!]
\centering
\caption{Top-pair cross sections computed at NNLO+NNLL accuracy with \toppp~v2.0, and product of branching ratios $\times$ acceptance $\times$ efficiency losses for the fully-hadronic and $\ell+$\,jets (with $\ell^\pm = e^\pm, \mu^\pm$) final states (for the selection criteria listed in Sections~\ref{subsec:ttbar_fulljets} and \ref{subsec:ttbar_ljets} respectively, computed with \MGaMC), in pp collisions at $\sqrts$~=~14 and 100~TeV. 
\label{tab:acceptance}}
\resizebox{\textwidth}{!}{%
\begin{tabular}{lc|cc}\toprule
\multirow{2}{*}{$\sqrt{s}$} & \multirow{2}{*}{$\sigma^{t\overline t,{\rm incl.}}_{\rm NNLO+NNLL}$} & \multicolumn{2}{c}{top-quark pair ($\mathcal{B}\,\times$ acceptance $\times$ efficiency) losses} \\
        &    &  fully hadronic &  $\ell+$\,jets \\\hline
 14 TeV & $980 \pm 17\,(\textsc{pdf})\,^{+24}_{-35} (\rm{scale})$~pb & $0.457 \times 0.57 \times 0.75 = 0.20$ & $0.35 \times 0.64 \times 0.75 = 0.17$ \\
100 TeV & $34.80 \pm 1.20\,(\textsc{pdf})\,^{+1.00}_{-1.65} (\rm{scale})$~nb & $0.457 \times 0.76 \times 0.75 = 0.26$ & $0.35 \times 0.71 \times 0.75 = 0.19$ \\
\bottomrule 
\end{tabular}
}
\end{table*}

\subsubsection{Fully hadronic  \texorpdfstring{$\ttbar$}{ttbar} final state\label{subsec:ttbar_fulljets}}

One can first attempt the $t\to {\overline B}^0_{(s)}+q$ measurement in the multi-quark decay of a $\ttbar$ pair: $\ttbar \to W^+\,b\,W^-\overline{b}\to q(\overline{q}'b)\,\, q''\overline{q}'''\overline{b}$, where both $W$ bosons decay hadronically (and, in our case of interest, one of its down-type quarks recombines with the closest $b$-quark to form a neutral $B$ meson). Such a fully-hadronic $\ttbar$ final state has the largest branching fraction ($\mathcal{B}=0.457$), and provides two top quarks per event, although it has the experimental drawback of a large combinatorial background from QCD multi-jets processes. Notwithstanding this difficulty, ATLAS and CMS have managed to setup online triggers that can collect all hadronic $\ttbar$ events without any level-1 prescale rate. Typical selection criteria used by the ATLAS/CMS experiments~\cite{Aad:2014zea,Aaboud:2017mae,Aaboud:2018eqg,Chatrchyan:2013xza,Khachatryan:2015fwh,Sirunyan:2018mlv} are:
\begin{itemize}
\item At least 4 jets (reconstructed with the anti-$k_{\rm T}$ algorithm~\cite{Cacciari:2008gp,Cacciari:2011ma} with distance parameter $R=0.5$) with $\pt > 25$~GeV and $|\eta|<3.0$ ($|\eta|<5.0$ at the FCC);
\item A total scalar sum of transverse energy in the reconstructed jets above roughly twice the top mass, $\HT > 350$~GeV;
\item At least 1 $b$ jet tagged with a typical 75\% efficiency;
\end{itemize}
Based on such an event signature, the HL-LHC studies indicate that unprescaled triggers can collect all relevant $\ttbar$ events at level-1 without any significant loss~\cite{Azzi:2017iwa} although, if needed, one can further require at higher-level trigger (HLT) the extra presence of a high-$\pt$ $\overline{B}^0$ meson, to reduce the collection rate of potentially large backgrounds. We generate $\ttbar$ events with \MGaMC~v2.6.6 (\MGshort\ henceforth)~\cite{Alwall:2014hca} using the {\tt NNPDF30\_nlo\_as\_0118} PDF~\cite{Ball:2014uwa}, implementing (at the parton level) the selection criteria listed above, and obtain yield losses from the combined acceptance\,$\times$\,efficiency\,$= 0.57\times 0.75 = 0.43$ and $0.76\times 0.75 = 0.57$ at the HL-LHC and FCC, respectively (Table~\ref{tab:acceptance}). Combining those values with the numbers~(\ref{eq:N_tt_LHC})--(\ref{eq:N_tt_FCC}) and the overall branching ratio\footnote{Note that we count two top quarks per event and that the fully hadronic $\ttbar$ branching fraction enters as a square-root, $\sqrt{0.457}$, given that one top decay to $\overline{B}^0_{(s)}+$\,jet is already included in the $2.5\cdot10^{5}$ and $6\cdot 10^{7}$ event counts.}, results in $N_{\rm t\to \overline{B}^0_{(s)}+j}^{\ttbar\;\rm full\;had.} = 2 \times 2.5\cdot10^{5}\times \sqrt{0.457} \times 0.43 \approx 1.5\cdot 10^5$, and $2\times 6\cdot 10^{7}\times \sqrt{0.457} \times 0.57 \approx 4.5\cdot 10^7$ top-quarks theoretically reconstructible in the (sum of) $t\to {\overline B}^0_{(s)}+q$ decay modes in fully-hadronic $\ttbar$ samples, 
at the HL-LHC and FCC, respectively.

\subsubsection{Lepton+jets \texorpdfstring{$\ttbar$}{ttbar} final state \label{subsec:ttbar_ljets}}

The ``cleanest'' channel to attempt the two-body top-decay measurement is the lepton+jets one, $\ttbar \to W^+\,b\,W^-\overline{b}\to q\,(\overline{q}'b)\,\, \ell\nu_\ell\overline{b}$, where $\ell$ stands for an electron or muon and, in our case of interest, one of the down-type quarks from the $W$ boson decay further recombines with the closest $b$-quark to form a neutral $B$ meson. As done commonly at the LHC, we exclude the channels with hadronically-decaying tau leptons, which are more difficult to reconstruct, but include electrons and muons from tau decays that have a combined top-quark branching fraction of $\mathcal{B} = 6.4\%$. Compared to the fully-hadronic mode, the $\ell$-plus-jets channel has a slightly smaller $\mathcal{B}\approx$~35\%  overall branching ratio, and half the number of top quarks available for the 2-body decay, but it features a cleaner experimental signature based on the presence of a high-$\pt$ isolated electron or muon. 
The following selection criteria, typically used by the ATLAS/CMS experiments~\cite{Chatrchyan:2011ew,Aad:2012qf,Aad:2015pga,Khachatryan:2016yzq,Sirunyan:2017uhy,Aaboud:2017cgs}, are applied:
\begin{itemize}
\item One isolated (cone radius $R_{\rm isol} = 0.3$) charged lepton $\ell$ with $\pt > 30$~GeV and $|\eta|<2.5$  ($|\eta|<5.0$ at the FCC);
\item At least 2 jets (reconstructed with the anti-$k_{\rm T}$ algorithm with $R=0.5$) with $\pt > 25$~GeV and $|\eta|<3.0$  ($|\eta|<5.0$ at the FCC), and separated from the lepton by $\Delta R (\ell,j) > 0.4$;
\item 1 $b$-jet tagged with a typical 75\% efficiency;
\end{itemize}
Based on such an event signature (one isolated high-$\pt$ charged lepton, plus at least 2 jets of which one is $b$-tagged), a trigger can be implemented to collect all relevant $\ttbar$ events unprescaled without any significant loss. 
[If needed to reduce any unforeseen extra background, one could also add as HLT requirements a minimum missing transverse energy from the unobserved $\nu$, and/or the extra presence of a high-$\pt$ $\overline{B}^0$ meson in the event.] The impact of such selection criteria, evaluated again at NLO accuracy with \MGshort\ at the parton level, indicates a 64\% (71\%) acceptance of the $\ell+$\,jets $\ttbar$ cross section at 14 (100) TeV (Table \ref{tab:acceptance}). These numbers are very similar to those obtained with the \mcfm\ v8.0 code~\cite{Campbell:2010ff}, with a very small dependence on the underlying PDF~\cite{dEnterria:2017dum}. The extra requirement to have one $b$-tagged jet in the event results in a final combined acceptance$\times$efficiency~$\approx$~48\% (53\%) to collect such a $\ttbar$-enriched sample at 14 (100) TeV. 
Thus, combining the numbers~(\ref{eq:N_tt_LHC})--(\ref{eq:N_tt_FCC}) with the overall $\ell$+jets branching ratio 
and the acceptance and efficiency losses, we expect about $N_{\rm t\to \overline{B}^0_{(s)}+j}^{\ttbar\; \ell+\rm jet} =~2.5\cdot10^{5} \times 0.35/\sqrt{0.457} \times 0.48 \approx 6.2\cdot 10^4$ and $6\cdot 10^{7} \times 0.35/\sqrt{0.457} \times 0.53 \approx 1.6\cdot 10^{7}$ top-quarks theoretically reconstructible in the (sum of) $t\to {\overline B}^0_{(s)}+q$ decay modes at the HL-LHC  and FCC, respectively.

\subsection{Measurement of the \texorpdfstring{$t\to {\overline B}^0_{(s)}+$\,jet}{t to B0(s)+jet} decay}

The last step needed to experimentally observe the rare top decays considered here relies on estimating the acceptance and efficiency to measure the decay ${\overline B}^0_{(s)}$ mesons, given that the reconstruction of the accompanying back-to-back ($u,c$) jet is already accounted for in the event count after the typical selection criteria discussed in the previous Section. For the full ${\overline B}^0_{(s)}$-meson reconstruction required in this work, one cannot use the $\overline{B}^0_{(s)}\to\ell\nu_\ell\,X$ (semi)leptonic decays with undetected neutrinos in the event, and must focus instead in the most abundant hadronic modes involving the $b \to c$ transition at the quark level either producing a charmonium- or charmed-meson plus lighter hadrons in the final state, $\overline{B}^0_{(s)}\to\jpsi\,h\,h$ (Table~\ref{tab:B0_jpsi_decays}) or $\overline{B}^0_{(s)}\to D\,h\,(h)$ (Table~\ref{tab:B0_D_decays}) with $h = \pi, K$. It is important to note that, for our signal events, the presence of a truly high-$\pt$ $\overline{B}^0_{(s)}$-meson ---carrying at least a momentum corresponding to half of the top-quark mass, $\pt^B \gtrsim 85$~GeV, \ie\ with about a factor of 10 larger transverse momenta than those typically studied ($\pt^B\gtrsim 5$~GeV) in more inclusive $B$-meson measurements at the LHC--- will significantly boost all final decay hadrons ($\jpsi$, $D$, $\pi$, and $K$), thereby improving the reconstruction performance. 
The detector acceptance for each final state can be realistically quantified as discussed below, although the full computation of $\mu$/$\pi$/$K$ reconstruction efficiencies would require a dedicated analysis of additional signal losses (due to tracking and secondary vertexing inefficiency, particle misidentification, combinatorial backgrounds, bin migrations, detector resolution effects, etc.) that go well beyond this first feasibility study. Since all LHC studies indicate that the reconstruction efficiency (in particular for tracking and vertexing) increases steadily with $\pt^B$ for all decay modes~\cite{Aad:2012jga,Aad:2012kba,Aad:2012pn,Aad:2015zix,Chatrchyan:2011pw,Chatrchyan:2011vh,Chatrchyan:2012rga,Chatrchyan:2013cda,CMS:2014xfa,Khachatryan:2015isa,Khachatryan:2015uja,Khachatryan:2016wqo,Sirunyan:2018zys,Sirunyan:2017xss,Sirunyan:2018ktu,LHCb:2011aa,Aaij:2011ac,LHCb:2012ae,Aaij:2013noa,Aaij:2013orb,Aaij:2013zpt,CMS:2014xfa,Aaij:2015fea,Aaij:2011jp,Aaij:2012zka,Aaij:2013dda,Aaij:2013pua,Aaij:2015sqa,Aaij:2018jqv,Aaij:2018rol,Aaij:2019lkm}, no extra reconstruction efficiency loss will be hereafter assumed in the estimation of final event yields.

\subsubsection{Final states with \texorpdfstring{$\jpsi$}{Jpsi} mesons}

The ATLAS~\cite{Aad:2012jga,Aad:2012kba,Aad:2012pn}, CMS~\cite{Chatrchyan:2011pw,Chatrchyan:2011vh,Chatrchyan:2012rga,Chatrchyan:2013cda,CMS:2014xfa,Khachatryan:2015isa,Khachatryan:2015uja,Khachatryan:2016wqo,Sirunyan:2018zys}, and LHCb~\cite{LHCb:2011aa,Aaij:2011ac,LHCb:2012ae,Aaij:2013noa,Aaij:2013orb,Aaij:2013zpt,CMS:2014xfa,Aaij:2015fea} experiments have measured neutral $B$ mesons in their clean exclusive decays into $\jpsi$, followed by its dimuon decay with $\mathcal{B}(\jpsi\to\mu^+\mu^-) = 5.96\%$, plus kaons and/or pions, with the branching ratios listed in Table~\ref{tab:B0_jpsi_decays}. The combined branching fractions for such final states are rather low, in the $10^{-4}$--$10^{-5}$ range, but relatively free of backgrounds.


\begin{table*}[htpb!]
\centering
\caption{Most important exclusive $\overline{B}^0$ and $\overline{B}^0_s$ decay modes into $\jpsi$+hadrons, with total branching ratios (product of each consecutive decay, from~\cite{PDG}), and typical acceptances for the corresponding final-state decay products in the considered $(\mu,\pi,K)$ central detector acceptances at the LHC and FCC (see text).
\label{tab:B0_jpsi_decays}}
\resizebox{\textwidth}{!}{%
\begin{tabular}{llcc}\toprule
 $\overline{B}^0_{(s)}$ meson & Total branching fraction & \multicolumn{2}{c}{Acceptance} \\
 decay mode & (product of individual $\mathcal{B}$'s) & LHC & FCC \\\hline
$\overline{B}^0\to\jpsi(\mu^+\mu^-)\,K^{*0}(K^\pm\pi^\mp)$ & $1.27\cdot 10^{-3}\times 0.0596 \times  1 \approx 7.6\cdot10^{-5}$ & 50\% & 55\% \\
$\overline{B}^0\to\jpsi(\mu^+\mu^-)\,K^+\pi^-$ & $1.15\cdot 10^{-3}\times 0.0596\approx 6.9\cdot 10^{-5}$ & 60\%  & 70\%  \\
$\overline{B}^0\to\jpsi(\mu^+\mu^-)\,K^0_s(\pi^+\pi^-)$ & $0.87\cdot 10^{-3}\times 0.0596 \times 0.69\approx 3.6\cdot 10^{-5}$ & 50\% & 60\%  \\\hline 
$\overline{B}^0_s\to\jpsi(\mu^+\mu^+)\,\phi(K^+K^-)$ & $1.08\cdot 10^{-3}\times 0.0596 \times 0.49 \approx 3.2\cdot 10^{-5}$ & 60\%  & 65\%  \\
$\overline{B}^0_s\to\jpsi(\mu^+\mu^+)\,K^+K^-$ & $0.79\cdot 10^{-3}\times 0.0596 \approx 4.7\cdot 10^{-5}$ & 65\%  & 70\%   \\
$\overline{B}^0_s\to\jpsi(\mu^+\mu^+)\,\pi^+\pi^-$ & $0.21\cdot 10^{-3}\times 0.0596 \approx 1.25\cdot 10^{-5}$ & 65\%  & 70\%  \\
\bottomrule
\end{tabular}
\label{tab:B0_Jpsi_decays}
}
\end{table*}

The typical reconstruction of the $\overline{B}^0_{(s)}$ events of interest proceeds as follows (see \eg~\cite{Khachatryan:2016wqo} for measurements of $B\to \jpsi+X$ decays in the $\ttbar$ lepton+jet final states of relevance here). The $\overline{B}^0_{(s)}$ candidates are obtained by: (i) reconstructing two muons within $|\eta|<3$~(5 at FCC) and matching their invariant mass to that of a non-prompt $\jpsi$, (ii) reconstructing two additional tracks over $|\eta|<3$~(5 at FCC) and requiring them (with charged pion and/or kaon mass hypothesis) to have an invariant mass consistent with any intermediate hadronic resonance ($K^{*}(892)^{0}$, $K^0_s$, $\phi$(1020),...) if present, (iii) fitting the charged tracks to a common secondary vertex consistent with the $\overline{B}^0_{(s)}$-meson lifetime, and finally (iv) requiring the $\jpsi$ and accompanying (resonant or not) hadronic state to have an invariant mass matching the $\overline{B}^0_{(s)}$ one. In order to assess the geometrical acceptance of all decay channels of Table~\ref{tab:B0_jpsi_decays}, we generate $\ttbar$ events with \pythia~8~v2.26  (\pythia~8, henceforth)~\cite{Sjostrand:2014zea}, selecting only those where top quarks decay into a $\overline{B}^0_{(s)}$ meson followed by the decay of the latter into each individual mode listed, switched-on one by one in the code. We then verify whether the two muons from the $\jpsi$ and the pions and/or kaons are all within the detector acceptance (we take $|\eta|<3,5$ for all tracks at the LHC and FCC, respectively, and require both muons to have $\pt> 3$~GeV). With such a Monte Carlo event generation setup, we obtain the acceptances listed in the last two columns of Table~\ref{tab:B0_jpsi_decays}, which lie in the 50\% to 70\% range.

\begin{table*}[htpb!]
\centering
\caption{Expected number of $t\to \overline{B}^{0}_{(s)}+$\,jet events in selected $\ttbar$ final-states (fully hadronic, and $\ell+$\,jets)  with $\overline{B}^0$-mesons decaying into $\jpsi$ plus light hadrons, after estimating branching ratios and detector acceptance losses, for pp collisions at the LHC and FCC.
\label{tab:B0_jpsi_decay_yields}}
\begin{tabular}{llcc}\toprule
Top pair    & $\overline{B}^0_{(s)}$ meson & \multicolumn{2}{c}{Events after all selection cuts} \\
final state & decay mode        &  LHC (3~ab$^{-1}$) & FCC (20~ab$^{-1}$) \\\hline
$\ttbar \to \overline{B}^{0}+u,c\,$\, & & & \\
\textopenbullet\; fully hadronic & \multirow{2}{*}{$\overline{B}^0\to\jpsi(\mu^+\mu^-)\,K^+\pi^-$} & $3.1$ & $1050$\\
\textopenbullet\; $\ell+$\,jets &  & $1.4$ & $380$ \\
\textopenbullet\; fully hadronic & \multirow{2}{*}{$\overline{B}^0\to\jpsi(\mu^+\mu^-)\,K^{*0}(K^\pm\pi^\mp)$} & $2.7$ & $940$\\
\textopenbullet\; $\ell+$\,jets & & $1.2$ & $340$ \\
\textopenbullet\; fully hadronic & \multirow{2}{*}{$\overline{B}^0\to\jpsi(\mu^+\mu^-)\,K^0_s(\pi^+\pi^-)$} & $1.3$ & $470$ \\
\textopenbullet\; $\ell+$\,jets &  & $0.6$ & $170$ \\
Sum of all channels &  & $10.2$ & $3\,300$ \\\hline

$\ttbar \to \overline{B}^{0}_{s}+c,u\,$\, & & & \\
\textopenbullet\; fully hadronic  & \multirow{2}{*}{$\overline{B}^0_s\to\jpsi(\mu^+\mu^+)\,K^+K^-$} & $2.1$ & $710$ \\
\textopenbullet\; $\ell+$\,jets &  & $0.9$ & $250$ \\
\textopenbullet\; fully hadronic  & \multirow{2}{*}{$\overline{B}^0_s\to\jpsi(\mu^+\mu^+)\,\phi(K^+K^-)$} & $1.4$ & $460$\\ 
\textopenbullet\; $\ell+$\,jets & & $0.6$ & $170$ \\ 
\textopenbullet\; fully hadronic & \multirow{2}{*}{$\overline{B}^0_s\to\jpsi(\mu^+\mu^+)\,\pi^+\pi^-$} & $0.6$ & $200$\\
\textopenbullet\; $\ell+$\,jets &  & $0.3$ & $70$\\
Sum of all channels & & $5.7$ & $1\,900$ \\\hline
TOTAL & & 16 & 5\,200 \\
\bottomrule
\end{tabular}
\end{table*}

Based on the analysis outlined above, Table~\ref{tab:B0_jpsi_decay_yields} lists the final expected number of two-body $t\to \overline{B}^{0}_{(s)}+$\,jet events in $\ttbar$ (fully hadronic, and $\ell+$\,jets) final states with the $\overline{B}^0_{(s)}$-meson decaying into $\jpsi$ plus light hadrons, after accounting for branching ratios and detector acceptance losses (Table~\ref{tab:B0_Jpsi_decays})
in pp collisions at the LHC and FCC. Among the $\overline{B}^{0}_{(s)}$-meson decay channels, the most promising ones in terms of final yields are $\overline{B}^0\to \jpsi\,K^+\pi^-$, $\overline{B}^0\to \jpsi\,K^{*0}$, and $\overline{B}^0_s\to\jpsi\,K^+K^-$.
We expect a total of about 16 and 5\,200 rare two-body top-quark signal events reconstructed at the LHC and FCC, respectively, with the final states with reconstructed $\overline{B}^0$ mesons being about twice more abundant than those with $\overline{B}^0_{s}$ mesons. The $\ttbar$ event selection based on fully-hadronic final-states leads to about a factor of two larger signal counts than that from $\ell+$\,jets. The final states with $c$- or $u$-quark jets, accompanying the $\overline{B}^0_{(s)}$ meson, share evenly the final number of signal events. The total of $\sim$16 signal events expected at the LHC indicate a difficult measurement of the 2-body top-quark decays, via an invariant mass analysis of $\overline{B}^0(\jpsi\,h\,h)$ channels plus a jet, and one will need a future hadron collider such as the FCC to carry out the measurement with about $\times$300 times more reconstructed events available.
 
\subsubsection{Final states with \texorpdfstring{$D^{0,\pm}$}{D} mesons}

The previous section considered the measurement of two-body ${\overline B}^0_{(s)}+$\,jet top-quark decays via $\overline{B}^0_{(s)}$ mesons reconstructed in their $\jpsi(\mu^+\mu^-)\,hh$ final states that feature very clean topologies albeit relatively small decay rates. Alternative $\overline{B}^0_{(s)}$ meson decays exist into multi-$\pi$,$K$ final states through an intermediate $D$ meson with larger branching ratios (for updated comprehensive ${\overline B}^0_{(s)}$ decay lists, see \eg Refs.~\cite{PDG,Amhis:2019ckw}). Potentially large combinatorial backgrounds for such channels can be reduced if one avoids final states with neutral hadrons and with more than four tracks, which can be more difficult to reconstruct under high pp pileup conditions. Namely, focusing on $\overline{B}^0_{(s)}\to D^{0,\pm}_{(s)}\,h(h)$ modes (with $h = \pi,K$) with $\mathcal{B} \approx 10^{-4}$--$10^{-3}$ followed by $\mathcal{B}(D^+\to K^-\pi^+\pi^+) = 0.094$,  $\mathcal{B}(D^+_s\to K^-K^+\pi^+) = 0.055$, or $\mathcal{B}(D^0\to K^-\pi^+) = 0.039$, with factors of 2--5 larger final branching fractions than the $\jpsi$-based ones discussed in the previous subsection. Among those, the decays listed in Table~\ref{tab:B0_D_decays} feature the largest total branching ratios.


\begin{table*}[htpb!]
\centering
\caption{Most important $\overline{B}^0$ and $\overline{B}^0_s$ decay modes into $D$-mesons plus one (or two) light mesons, with total branching ratios (product of each consecutive decay, from~\cite{PDG}), and typical acceptances for the corresponding final-state decay products in the considered $(\pi,K)$ central detector acceptances at the LHC and FCC.
\label{tab:B0_D_decays}}
\resizebox{\textwidth}{!}{%
\begin{tabular}{llcc}\toprule
 $\overline{B}^0_{(s)}$ meson & Total branching fraction & \multicolumn{2}{c}{Acceptance} \\
 decay mode & (product of individual $\mathcal{B}$'s) & LHC & FCC \\\hline
$\overline{B}^0\to D^+(K^-\pi^+\pi^+)\;\pi^-$ &  $2.52\cdot 10^{-3}\times 0.094\approx 2.37\cdot 10^{-4}$ & $55\%$  & $65\%$ \\
$\overline{B}^0\to \overline{D}^{*,+}(D^0(K^-\pi^+)\pi^+)\;\pi^-$ &  $2.74\cdot 10^{-3}\times 0.677\times 0.039\approx 7.2\cdot 10^{-5}$ & $55\%$  & $65\%$ \\
$\overline{B}^0\to \overline{D}^{0}(K^-\pi^+)\;\pi^+\pi^-$ &  $0.88\cdot 10^{-3}\times 0.039\approx 3.43\cdot 10^{-5}$ & $75\%$  & $80\%$ \\
$\overline{B}^0\to D^+(K^-\pi^+\pi^+)\;K^-$ & $0.186\cdot 10^{-3}\times 0.094\approx 1.75\cdot 10^{-5}$ & $55\%$  & $65\%$  \\\hline
$\overline{B}^0_s\to D_s^{+}(K^-K^+\pi^+) \pi^-$ & $3.0\cdot 10^{-3}\times 0.055 \approx 1.6\cdot 10^{-4}$ & $60\%$  & $60\%$ \\
$\overline{B}^0_s\to \overline{D}^{0}(K^-\pi^+)\;K^-\pi^+$ & $1.04\cdot 10^{-3}\times 0.039 \approx 4\cdot 10^{-5}$ & $60\%$  & $65\%$  \\
\bottomrule
\end{tabular}
}
\end{table*}

All experiments at the LHC (ATLAS~\cite{Aad:2015zix}, CMS~\cite{Khachatryan:2016wqo,Sirunyan:2017xss,Sirunyan:2018ktu}, and LHCb~\cite{Aaij:2011jp,Aaij:2012zka,Aaij:2013dda,Aaij:2013pua,Aaij:2015sqa,Aaij:2018jqv,Aaij:2018rol,Aaij:2019lkm}) have reconstructed\footnote{In particular, Ref.~\cite{Khachatryan:2016wqo} has explicitly measured $B\to D$ meson decays in the $\ttbar$ lepton+jet final states of relevance here.} the different ${\overline B}^0 \to D \to 3\pi\,K,\,2\pi\,2K$ decay chains listed in Table~\ref{tab:B0_D_decays}. Such analyses usually proceed as follows: (i) reconstructing two or three charged-hadron tracks within $|\eta|<3$~(5 at FCC) successfully fitted to a common vertex consistent with the $D$-meson lifetime, (ii) assigning to the tracks the pion or kaon masses, based on appropriate charge associations, and requiring the $m_{K\pi(\pi)}$ invariant mass to be consistent with the $D^{0,\pm}_{(s)}$ mass, and (iii) reconstructing one or two more charged-hadron tracks within $|\eta|<3$~(5 at FCC) that, combined with the $D^{0,\pm}_{(s)}$ candidate at a secondary vertex consistent with the $\overline{B}^{0}_{(s)}$-meson lifetime, have an invariant mass matching the $\overline{B}^0_{(s)}$ one. As done in the previous section, the geometric acceptance of all individual modes listed in Table~\ref{tab:B0_D_decays} is determined with $\ttbar$ events generated with \pythia~8, where the aforementioned analysis chain is implemented. Based on such a simulation setup, we obtain the acceptances listed in the last two columns of Table~\ref{tab:B0_D_decays}, which lie in the 55\% to 80\% range.

\begin{table*}[htpb!]
\centering
\caption{Expected number of $t\to \overline{B}^{0}_{(s)}+$\,jet events in selected $\ttbar$ final-states (fully hadronic, and $\ell+$\,jets) with $\overline{B}^0_{(s)}$-meson decays into a $D$-meson plus 1 or 2 light mesons, after estimating branching ratios and detector acceptance losses, in pp collisions at the LHC and FCC. \label{tab:B0_D_decays_yields}}
\begin{tabular}{llcc}\toprule 
Top pair & $\overline{B}^0_{(s)}$ meson & \multicolumn{2}{c}{Events after all selection cuts} \\
final state & decay mode        &  LHC (3~ab$^{-1}$) & FCC (20~ab$^{-1}$) \\\hline
$\ttbar \to \overline{B}^{0}+u,c\,$\, & & & \\
\textopenbullet\; fully hadronic  & \multirow{2}{*}{$\overline{B}^0\to D^+(K^-\pi^+\pi^+)\;\pi^-$} & $9.3$ & $3\,500$\\
\textopenbullet\; $\ell+$\,jets&  & $4$ & $1\,200$\\
\textopenbullet\; fully hadronic & \multirow{2}{*}{$\overline{B}^0\to \overline{D}^{*,+}(D^0(K^-\pi^+)\pi^+)\;\pi^-$} & $2.8$ & $1\,100$ \\ 
\textopenbullet\; $\ell+$\,jets & & $1.2$ & $380$ \\ 
\textopenbullet\; fully hadronic & \multirow{2}{*}{$\overline{B}^0\to \overline{D}^{0}(K^-\pi^+)\pi^+\;\pi^-$} & $1.8$ & $630$ \\
\textopenbullet\; $\ell+$\,jets&  & $0.8$ & $220$ \\
\textopenbullet\; fully hadronic  & \multirow{2}{*}{$\overline{B}^0\to D^+(K^-\pi^+\pi^+)\;K^-$} & $0.7$ & $250$\\
\textopenbullet\; $\ell+$\,jets&  & $0.3$ & $90$ \\
Sum of all channels  &  & $21.0$ & $7\,300$\\\hline

$\ttbar \to \overline{B}^{0}_{s}+c,u\,$\, & & & \\
\textopenbullet\; fully hadronic & \multirow{2}{*}{$\overline{B}^0_s\to D_s^{+}(K^+K^-\pi^+) \pi^-$} & $7.1$ & $2\,200$ \\ 
\textopenbullet\; $\ell+$\,jets &  & $3.0$ & $800$ \\ 
\textopenbullet\; fully hadronic  & \multirow{2}{*}{$\overline{B}^0_s\to \overline{D}^{0}(K^-\pi^+)\;K^-\pi^+$} & $1.6$ & $570$ \\
\textopenbullet\; $\ell+$\,jets &  & $0.7$ & $200$\\
Sum of all channels  &  & $12.4$ & $3\,800$ \\\hline
TOTAL & & $33$ & $11\,000$\\
\bottomrule
\end{tabular}
\end{table*}

Following the method outlined above, Table~\ref{tab:B0_D_decays_yields} lists the total expected number of two-body $t\to \overline{B}^{0}_{(s)}+$\,jet events in $\ttbar$ (fully hadronic, and $\ell+$\,jets) final states with $\overline{B}^0_{(s)}$-mesons decaying into $D^{0,\pm}$ plus light hadrons, after accounting for branching ratios and detector acceptance losses (Table~\ref{tab:B0_D_decays})
for pp collisions at the LHC and FCC. As for the $\jpsi$ final states considered in the previous section, full reconstruction efficiency for the boosted final hadrons from high-$\pt$ ${\overline B}^0_{(s)}$ decays is assumed in the estimation of final event yields. Among the $\overline{B}^{0}_{(s)}$-meson decay channels, the most promising ones in terms of final yields are $\overline{B}^0\to D^+\,\pi^-$, and $\overline{B}^0_s\to D_s^{+}\,\pi^-$. In total, we expect about 33 and 11\,000 rare two-body top-quark signal events at the LHC and FCC, respectively, with about twice more final states reconstructed with $\overline{B}^0$ than $\overline{B}^0_{s}$ mesons. 
The final signal yields from the $\ttbar$ event selection based on fully-hadronic final-states is about twice larger than that from $\ell+$\,jets. Signal events are equally shared between those having a $c$- or $u$-quark jet plus the $\overline{B}^0_{(s)}$ meson.
With about 33 signal counts expected at the LHC, the observation of the 2-body top-quark decays via exclusive final states involving $\overline{B}^0_{(s)}\to D\,h(h)$ channels may be feasible, but such a measurement will definitely profit from the 11\,000 reconstructed events expected at the FCC. In the next section, we provide an alternative and simpler method to observe the rare 2-body top-quark decays with much larger data samples at both the LHC and FCC.

\section{Measurement of \texorpdfstring{$t\to b(\mathrm{jet})+c(\mathrm{jet})$}{t to b(jet)+c(jet)} in pp collisions at the LHC and FCC}
\label{sec:t_bc}

Given the large signal reduction factors involved in the exclusive ${\overline B}^0_{(s)}$ hadronic final states discussed in the previous section, we can try an observation of the two-body $t \to \overline{B}^{0}_{(s)}+c$ decay minimizing the $\overline{B}^{0}_{(s)}$ decay branching fraction losses, by using events with fully-reconstructed bottom and charm jets. Namely, by reconstructing the dijet final state $t \to j\rm{(charm)}+j$(bottom), where the bottom jet is obtained clustering all products from the $\overline{B}^{0}_{(s)}$ decay (plus any other low-$\pt$ hadron from any soft gluon potentially emitted previously from the parent $b$ quark), and where both heavy-quark jets have an invariant $m_{b,c}$ mass consistent with $\mt$. Such a final state is particularly attractive as it represents the decay of a top quark into two jets corresponding to the consecutively lighter (bottom and charm) quark families.
The alternative bottom--plus--light-quark dijet final state, from the $t \to \overline{B}^{0}_{(s)}+u$ decay, should be much more swamped by the large QCD multijet background and will not be further considered. To reduce multijet combinatorial backgrounds, the analysis is carried out in $\ttbar$ lepton+jets events that pass the trigger selection discussed in Section~\ref{subsec:ttbar_ljets}. 
In more detail, the final state of interest for this search is an event with one bottom and one charm jet (from the 2-body top decay of concern), plus an extra bottom jet and a high-$\pt$ isolated lepton from the decay of the other accompanying top quark. The following baseline event selection is considered. Events are required to have:
\begin{enumerate}
\item Exactly 3 jets tagged as coming from two bottom and one charm quarks respectively (with the tagging performances discussed below), with at least one $b$- plus $c$-jet pair having an invariant mass $120 \leq m_{b,c}/{\rm GeV} \leq 220$ and azimuthal separation $\Delta \phi(b,c)>0.5\pi$. Jets are reconstructed with the anti-$k_T$ algorithm with $R=0.5$, $\pt >25$ GeV, and $|\eta|<3.0$ (5.0) at 14 (100) TeV. In order to suppress backgrounds, and following the decay kinematics properties of the signal events discussed at the beginning of Section~\ref{sec:theory}, we further require the tagged $c$-jet to be boosted ($p_T> \mt/2\approx 80$ GeV). 
\item An isolated ($R_{\rm isol}=0.3$) charged lepton $\ell$ is required with $\pt>30$ GeV and $|\eta|<2.5$ (5.0) at 14 (100) TeV. Isolation is defined at the particle level, by requiring that the scalar sum of the $\pt$ of all charged and neutral particles within $R_{\rm isol}$ (except the charged lepton and any neutrinos) is less than 15\% of the muon or electron $\pt$.
\end{enumerate}
The following bottom- and charm-jets reconstruction performances are assumed based on existing heavy-flavour studies in $\ttbar$ events at the LHC~\cite{Sirunyan:2017ezt}:
\begin{itemize}
\item Bottom-quark jets: $b$-jet tagging efficiency: 75\%, $b$-jet mistagging probability for a $c$-quark: 5\%, and $b$-jet mistagging probability for $udsg$ (light quarks or gluon): 0.5\%.
\item Charm-quark jets: Two different $c$-jet working points are considered: (i) $c$-jet tagging efficiency: 65\%, $c$-jet mistagging probability for a $b$ quark: 10\%, and $c$-jet mistagging probability for $udsg$: 10\% [labeled `HcT' hereafter]; and (ii) $c$-jet tagging efficiency: 35\%, $c$-jet mistagging probability for a $b$-quark: 5\%, and $c$-jet mistagging probability for $udsg$: 1\% ['LcT' label].
\end{itemize}


The analysis criteria listed above are implemented into a NLO \MGshort\ simulation using the {\tt NNPDF30\_nlo\_as\_0118} PDF and \pythia~8 for the parton shower (PS) and hadronization, with the PS matched to the matrix elements via the MC@NLO prescription~\cite{Frixione:2002ik}.
The underlying event in pp collisions, of relevance for a realistic application of the lepton isolation criterion listed above, is generated with the 2013 Monash tune~\cite{Skands:2014pea}. The signal NLO cross sections are scaled to NNLO+NNLL accuracy by multiplying them by a $K = 1.2$ factor~\cite{Czakon:2013goa}.
Background samples are simulated with the same setup for the following processes that can produce similar final states: standard-decay $\ttbar$; single top ($t+W$, $t+j$ and $t+b$); $W+$jets ($W+udsg$, $W+c$, $W+b$) and $Z+$jets ($Z+g$, $Z+c$, $Z+b$) collectively labeled $V+\,$jet; dibosons ($VV = WW$, $ZZ$, $WZ$); and QCD multijets ($\bbbar+X$, $\ccbar+X$, $\qqbar+X$)\footnote{Full-NLO+PS multijet events are very heavy to produce in terms of computing time, and are generated here based on LO+PS simulations alone.}.\\

\begin{table*}[htpb!]
\centering
\caption{Total number of events expected for $t\bar{t}\to N_j+\ell^{\pm}+X$ with $\ell^\pm=\mu^\pm,e^\pm$, in pp collisions at $\sqrts$~=~14 and 100~TeV, passing the analysis criteria for the `HcT' and `LcT' charm jet reconstruction performances discussed in the text. The ``trigger" row  lists the number of events that pass the trigger cuts presented in subsection~\ref{subsec:ttbar_ljets}, and $N_{j_b}=2\,\&\&\,N_{j_c}=1$ indicates the exact requirement of two $b$- and one $c$-quark jets reconstructed in the event with the kinematic criteria discussed in the text.
\label{tab:cutflow}}
\resizebox{\textwidth}{!}{%
\begin{tabular}{l|cc|cc}\toprule
\multirow{2}{*}{Selection criteria} &  \multicolumn{2}{c|}{pp at $\sqrts = 14$ TeV (3~ab$^{-1}$)} &  \multicolumn{2}{c}{pp at $\sqrts = 100$ TeV (20~ab$^{-1}$)} \\
& `HcT' & `LcT' & `HcT' & `LcT' \\\hline
(0): None & $1.0\cdot 10^9$ & $1.0\cdot 10^9$ & $2.4\cdot 10^{11}$ & $2.4\cdot 10^{11}$ \\
(1): Trigger ($\ell+$jets) & $4.8\cdot 10^8$ & $4.8\cdot 10^8$ & $1.3\cdot 10^{11}$ & $1.3\cdot 10^{11}$ \\
(2): (1) $+$ $(N_{j_b}=2\,\&\&\,N_{j_c}=1)$ & $8.2\cdot 10^7$ & $3.8\cdot 10^7$ & $2.3\cdot 10^{10}$ & $1.2\cdot 10^{10}$\\
(3): (1) $+$ (2) $+$ $120 \leq m_{b,c}\leq 220$ GeV & $5.3\cdot 10^7$ & $2.6\cdot 10^7$ & $1.3\cdot 10^{10}$ & $7.3\cdot 10^9$\\
\bottomrule
\end{tabular}
}
\end{table*}

After full jet reconstruction and implementing the bottom and charm jet (mis)identification performances and basic cuts listed above, we obtain the $\ttbar$ yields listed in Table~\ref{tab:cutflow} after each one of the analysis steps. The final numbers after cuts are listed in Table~\ref{tab:t_c_b_jets} for signal and each one of the individual backgrounds. 

\begin{table*}[htpb!]
\centering
\caption{Total number of bottom+charm jet pairs (events, in parentheses) expected over $m_{b,c} = 120\mathrm{-}220$~GeV for signal and backgrounds after all analysis cuts (see text), in lepton+jets $\ttbar$ final states simulated in pp collisions at $\sqrts$~=~14 and 100~TeV, for the `HcT' and `LcT' charm jet reconstruction performances described in the text. The last row lists the expected signal significance $\pazocal{S}$ (in standard deviations) for the top-quark 2-body decay reconstructed as $t\to b$-jet+$c$-jet.
\label{tab:t_c_b_jets}}
\begin{tabular}{l|cc|cc}\toprule
\multirow{2}{*}{process} &  \multicolumn{2}{c|}{pp at $\sqrts = 14$ TeV (3~ab$^{-1}$)} &  \multicolumn{2}{c}{pp at $\sqrts = 100$ TeV (20~ab$^{-1}$)} \\
& `HcT' & `LcT' & `HcT' & `LcT' \\\hline
$\ttbar$ (signal) & 5\,300 & 2\,500 & $1.4\cdot10^6$ & $7.5\cdot10^5$ \\
$\ttbar$ (backgd.) & $2.1(1.7)\cdot 10^6$ & $9.3(7.4)\cdot 10^5$ & $3.0(2.4)\cdot 10^8$ & $1.1(0.9)\cdot 10^8$ \\
$tW$ & $7.0(5.8)\cdot 10^4$ & $3.1(2.5)\cdot 10^4$ & $1.2(1.0)\cdot 10^7$ & $4.7(4.4)\cdot 10^6$\\
$tj$ & $7.8(6.6)\cdot 10^4$ & $2.7(2.4)\cdot 10^4$ & $1.1(1.0)\cdot 10^7$ & $4.0(3.7)\cdot 10^6$\\
$V$+jet & $1.5(1.3)\cdot 10^5$ & $7.6(7.6)\cdot 10^4$ & $1.1(0.8)\cdot 10^7$ & $1.1(0.8)\cdot 10^7$\\
$VV$ & $7.7(6.3)\cdot 10^3$ & $2.8(2.3)\cdot 10^3$ & $9.1(7.3)\cdot 10^5$ & $3.4(3.1)\cdot 10^5$\\
Multijet & $<4.8\cdot 10^4$ & $<4.8\cdot 10^4$ & $<1.2\cdot 10^7$ & $<1.2\cdot 10^7$\\\hline
$\pazocal{S}$ (std.\,dev.) & 6.1 & 4.5 & 130. & 110. \\
\bottomrule
\end{tabular}
\end{table*}

\begin{figure}[htbp!]
\centering
\includegraphics[width=0.495\textwidth]{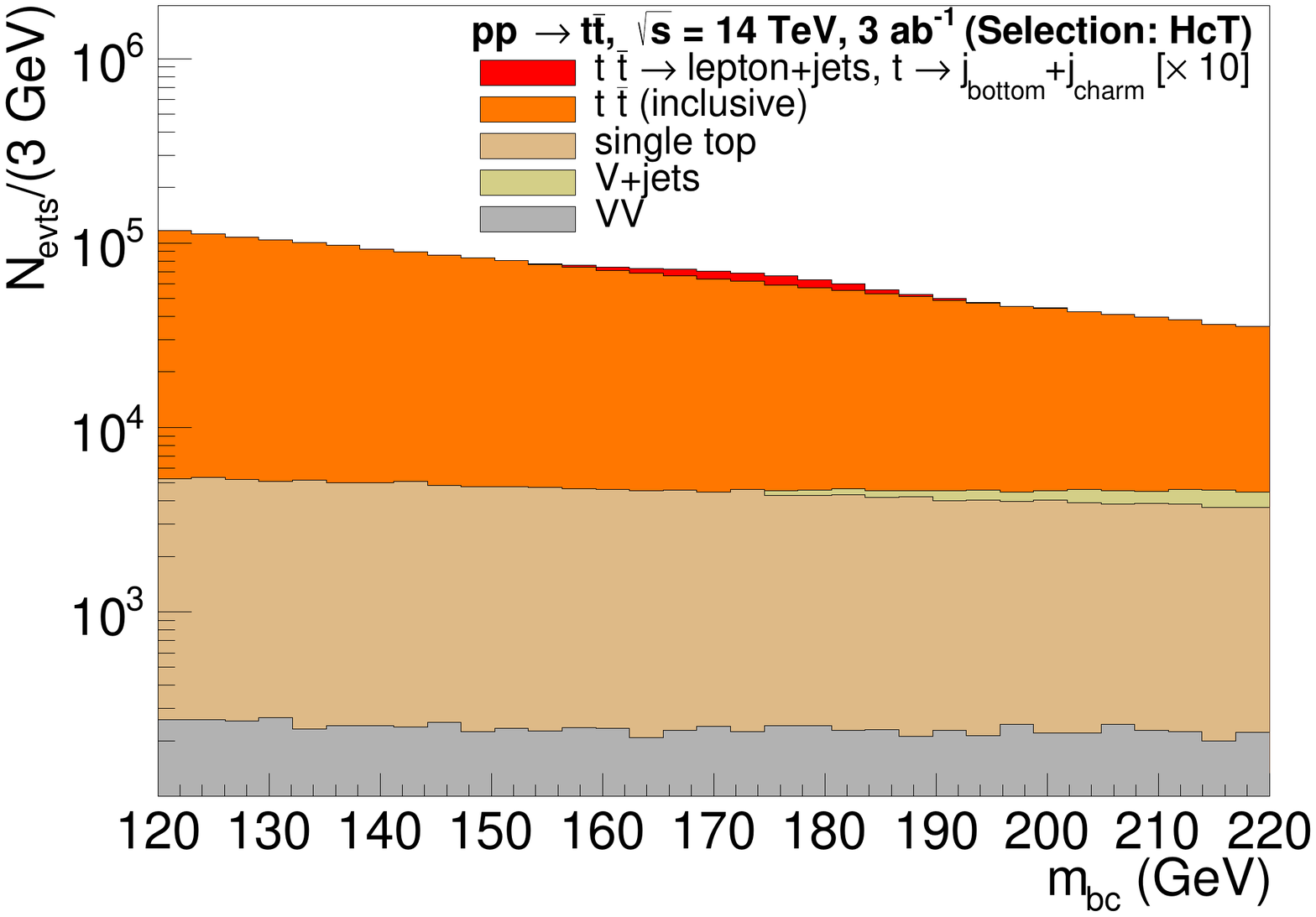}
\includegraphics[width=0.495\textwidth]{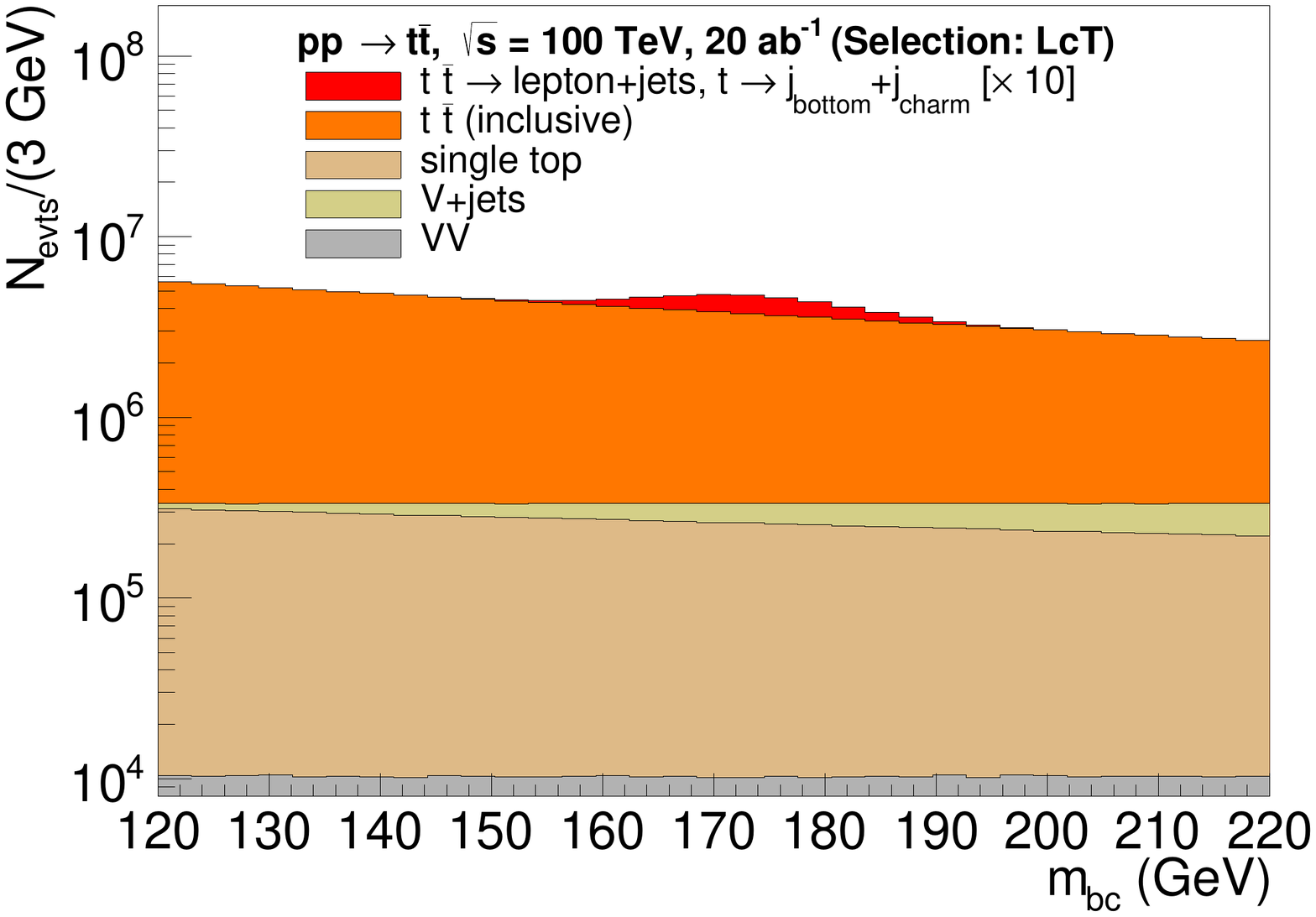}
\caption{Expected invariant mass distributions for pairs of charm and bottom jets reconstructed passing the analysis criteria for the $t\to b\,c$ analysis simulated for signal and backgrounds in pp collisions at $\sqrts$~=~14~TeV (left) and 100~TeV (right) for the `HcT' and `LcT' charm jet reconstruction performances, respectively, described in the text. The different histograms show the Gaussian signal (red, scaled by a factor of 10 for visibility) and individual (exponential-like) backgrounds, stacked one above each other, listed in Table~\ref{tab:t_c_b_jets}.
\label{fig:top_inv_mass}}
\end{figure}

The expected number of signal $\ttbar$ lepton+jets events where one top quark decays into a reconstructed pair of bottom and charm jets after cuts is $N_{S} = 5\,300$ and 1.4 million in pp collisions at the LHC and FCC, respectively, for the `HcT' selection. The numbers for the more stringent `LcT' charm reconstruction performance are about twice smaller. The applied selection criteria reduce the produced number of $t \to \overline{B}^{0}_{(s)}+c$ signal events by about a factor of six (ten) at the LHC (FCC), while reducing the backgrounds by two to three orders of magnitude. The most important remaining background passing the analysis cuts is that from $\ttbar$ production with standard top-quarks decays, which have a final expected number of events of about $N_{B} = 0.7$--1.7 and 90--240 millions at the LHC and FCC respectively, that are about one order of magnitude larger than the sum of the rest of backgrounds\footnote{The simulation of the QCD multijet backgrounds is very time-consuming, and only an upper limit in the number of events passing the analysis cuts can be provided from the zero events selected out of 10 million generated. In the real data analysis, those events will be well understood using control regions with appropriately reversed (``anti-signal'') cuts.}. Representative dijet mass distributions are shown in Fig.~\ref{fig:top_inv_mass} left (right) after all cuts, for the `HcT' (`LcT') selections at the LHC (FCC). 
The signal statistical significances are computed from the number of signal pairs over the square-root of the sum of all backgrounds pairs, $\pazocal{S} = N_{S}/\sqrt{N_{B}}$, in a window of 1.4 times the width of the reconstructed top mass, where we assume a dijet invariant mass resolution of $\delta m_{b,c} = 9$~GeV, which thereby translates into a significance calculated over the $m_{b,c} = 160\mathrm{-}190$~GeV mass range. No uncertainty in the background is assumed, as the continuum can be very accurately measured and fitted to a smooth function from the measured off-peak dijet mass distributions. The observation of the top-quark 2-body decay in the bottom+charm dijet final state is warranted both at the HL-LHC and FCC, as indicated by the $\pazocal{S} > 5$ std.\,dev.\ results listed in the last row of Table~\ref{tab:t_c_b_jets}. More advanced profile-likelihood statistical analyses for signal and backgrounds, typical of standard LHC analyses that go beyond the scope of this first study, can easily improve the simple significance estimates presented here that should thereby be considered as conservative lower limits.


\subsection{Top-quark mass via bottom+charm dijet decays}

\begin{figure}[htbp!]
\centering
\includegraphics[width=0.49\textwidth]{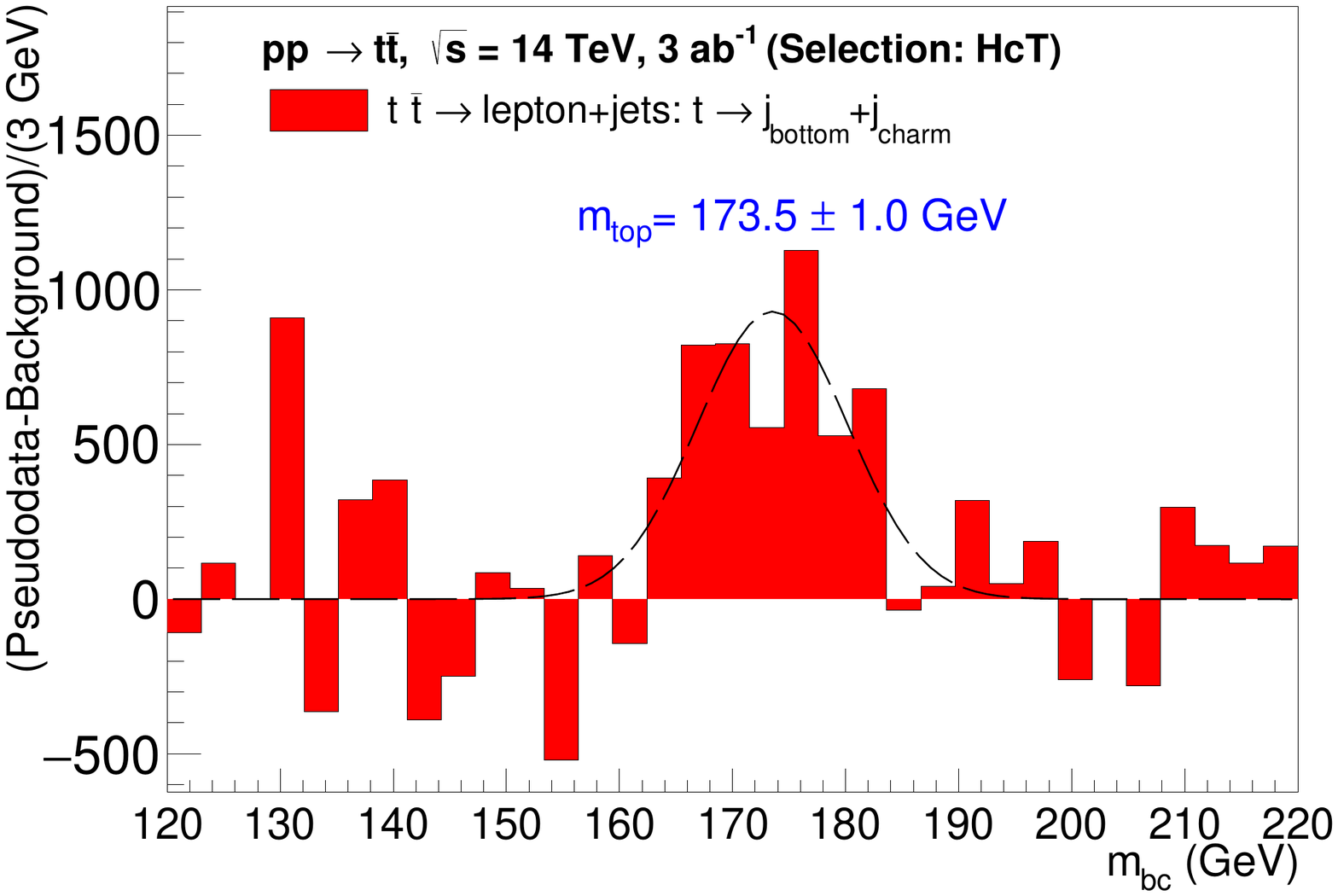}
\includegraphics[width=0.50\textwidth]{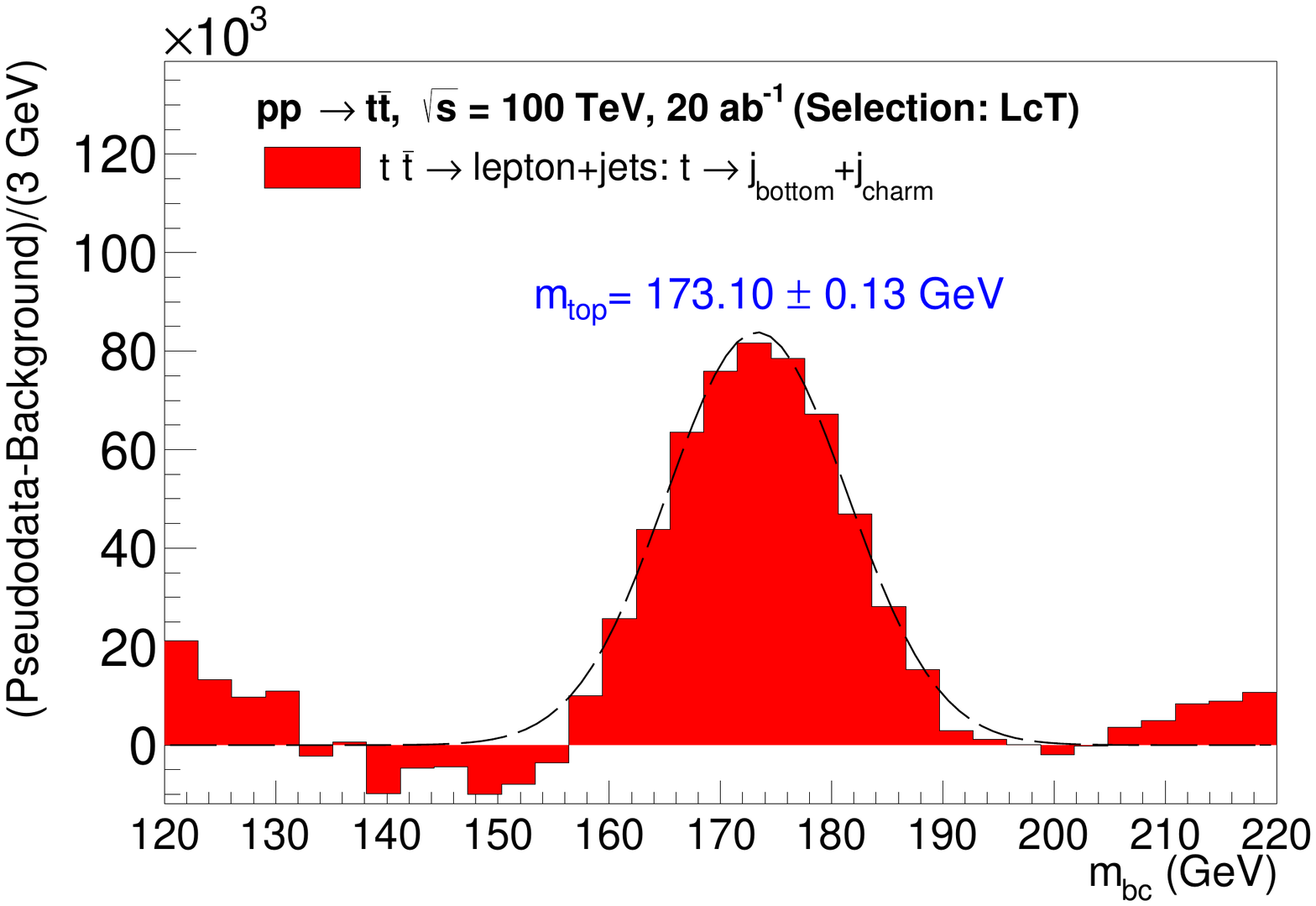}
\caption{Expected background-subtracted invariant mass distributions for pairs of charm and bottom jets reconstructed in the $t\to b\,c$ analysis in simulated pp collisions at $\sqrts$~=~14~TeV (left) and 100~TeV (right) after cuts for the `HcT' and `LcT' charm jet reconstruction performances, respectively, discussed in the text. The dashed line shows a Gaussian fit to the distributions, with the fitted value of the top-quark mass listed.\label{fig:top_mass}}
\end{figure}

The expected bottom-charm dijet invariant mass distributions after analysis cuts plotted in Fig.~\ref{fig:top_inv_mass} show a Gaussian-like signal peak, with a width driven by the experimental $\delta m_{b,c} = 9$~GeV dijet mass resolution assumed here, on top of the sum of all continuum backgrounds with an overall exponential- or powerlaw-like smooth shape. By fitting the $m_{b,c}$ distribution to an exponential-type function outside of the peak, one can estimate the expected background contributions below the top mass peak. The resulting 2-body top-quark signal mass distributions, with the fitted background subtracted, are plotted in Fig.~\ref{fig:top_mass} for pp collisions at $\sqrts$~=~14~TeV (left) and 100~TeV (right) for the `HcT' and `LcT' charm jet reconstruction performances, respectively. The values of the reconstructed top-quark mass and its uncertainty quoted in the plots are obtained from the corresponding Gaussian fit (dashed line).
We expect simple fit uncertainties in the top-quark mass of the order of 1~GeV at the LHC, and 0.15~GeV at the FCC. Beyond such uncertainties, statistical in nature, one would need to consider systematic uncertainties from the jet energy scale calibration that propagate into the dijet mass peak position. Nonetheless, with millions heavy-flavour jets reconstructed in dedicated studies in this energy range, the latter uncertainties should be well under control (and, in any case, would be smaller than in the standard $t\to b\qbar q'$ triple-jet final state). More advanced signal+background fit procedures exist, standard now in similar analyses at the LHC, that would provide a best-fit top-quark mass with potentially smaller uncertainties than those quoted here. We leave this more accurate work for an upcoming study. The top-quark mass determination method presented here has systematic uncertainties different (and, in some cases, smaller) than other extractions considered so far. Combining the top-quark mass measurement from the two-body charm+bottom dijet final state with those derived from other methods would thereby allow for a more accurate and precise extraction of this key SM parameter.


\section{Summary}
\label{sec:summary}

Rare two-body decays of the top quark into a neutral $B$-meson plus an up- or charm-quark jet, $t\to {\overline B}^0+u, c$ and $t\to {\overline B}^0_{s}+ c,u$, as well as into a bottomonium meson plus a jet $t \to \Upsilon(nS)+\, c,u$, have been studied for the first time. The leading-order (LO) branching ratios have been computed in a non-relativistic quantum chromodynamics (NRQCD) quark recombination model, and amount to $\mathcal{B}(t \to {\overline B}^0_{(s)}+\, {\rm jet}) \approx 4\cdot 10^{-5}$, and $\mathcal{B}(t \to \Upsilon(nS)+\, {\rm jet}) \approx 2.5\cdot 10^{-9}$. The feasibility of an experimental observation of such decays in multiple final states has been studied in $\ttbar$ pair production events in proton-proton collisions at $\sqrts = 14$~TeV at the HL-LHC and at 100~TeV at the FCC with integrated luminosities of 3~ab$^{-1}$ and 20~ab$^{-1}$ respectively. For each one of the final states considered, we take into account realistic estimates for their acceptance and reconstruction efficiencies based on current LHC studies. Table~\ref{tab:final_yields} summarizes the expected number of events in various decay final-state channels considered. Combining the results from all exclusive ${\overline B}^0_{(s)}$ decays, one expects a grand total of about 50 (16\,000) two-body top-quark decays reconstructed at the LHC (FCC) in the $\jpsi\,hh$ and $D^{0,\pm}\,h(h)$ final states (with $h=\pi,K$), respectively.
The expected numbers of top-quark 2-body decays, $t\to \overline{B}^0_{(s)}+c$, where the final state is reconstructed as a bottom-plus-charm dijet system are 5\,300 and 1.4 million after cuts in pp collisions at the LHC and FCC, respectively. A clear observation above backgrounds of the top-quark 2-body decay in such intriguing bottom+charm dijet final state appears warranted both at the LHC and FCC.


\begin{table*}[htpb!]
\centering
\caption{Summary of the total number of top-quark events expected in the various final states considered in this work for pp collisions at $\sqrts$~=~14 TeV (3~ab$^{-1}$) and 100~TeV (20~ab$^{-1}$). The first two columns list the number of produced pair $\ttbar$ and two-body $t$-quark decay events expected. The last three columns list the number of 2-body $\ttbar\to ({\overline B}^0_{s}+ c,u) + X$ top-quarks events expected in measurements using 
${\overline B}^0_{(s)}\to\jpsi\,hh$ and ${\overline B}^0_{(s)}\to D^{0,\pm}\,h(h)$ with $h=\pi,K$, and $t\to j_b+j_c$ final states, after accounting for their respective branching fractions, acceptance, and reconstruction efficiencies.\label{tab:final_yields}}
\resizebox{\textwidth}{!}{%
\begin{tabular}{l|cc|ccc}\toprule
pp system  & \multicolumn{2}{c}{Number of produced} & \multicolumn{3}{c}{Number of 2-body top-quark events}\\
$\sqrts,\; \LumiInt$ & \multicolumn{2}{c}{events expected} & \multicolumn{3}{c}{after $\mathcal{B}\,\times$ acceptance $\times$ efficiency cuts}\\
 & $\ttbar$ & $t \to \overline{B}^{0}_{(s)}+$jet & $({\overline B}^0_{(s)}\to\jpsi\,hh)$ & $({\overline B}^0_{(s)}\to D^{0,\pm}_{(s)}\,h(h)$ & $t \to b + c$ \\\hline
14 TeV, 3 ab$^{-1}$  & $3\cdot10^{9}$  & $2.5\cdot10^{5}$ & 16 & 33 & 5\,300\\
100 TeV, 20 ab$^{-1}$ & $7\cdot10^{11}$ & $6\cdot10^{7}$ & 5\,200 & 11\,000 & $1.4\cdot 10^6$ \\
\bottomrule
\end{tabular}
}
\end{table*}

We have also estimated the possibility to use the $t\to j_b+j_c$ final state to determine the top-quark mass via a simple dijet invariant mass analysis. Gaussian fit mass uncertainties of the order of 1~GeV at the LHC and 0.15~GeV at the FCC are expected. Such a top-quark mass determination, with systematic uncertainties smaller and different than other extractions considered so far, would allow for a more accurate and precise extraction of this key parameter of the Standard Model.

\begin{acknowledgments}
We are thankful to Pedro da Silva and Adish Vartak for helpful discussions on top-quark measurements and dijet resonances fitting procedures at the LHC, respectively, and to Michelangelo Mangano for useful comments. The work of HSS is supported by
the ILP Labex (ANR-11-IDEX-0004-02, ANR-10-LABX-63).
\end{acknowledgments}



\bibliographystyle{JHEP}        
\bibliography{top_decay}



\end{document}